\documentclass[%
 prx,
 amsmath,amssymb,
preprint,%
]{revtex4-1}

\usepackage{graphicx}
\usepackage{dcolumn}
\usepackage{bm}
\usepackage{bbm}
\usepackage{xcolor}
\usepackage{tikz}
\usetikzlibrary{positioning}
\usepackage{array}
\usepackage{mathtools} 
\usepackage{extarrows}

\usepackage{amsmath}
\usepackage{qcircuit}

\begin{document}
\title{Quantum optimal control with quantum computers: an hybrid algorithm featuring machine learning optimization.}

\author{Davide Castaldo}
\affiliation{ 
Dipartimento di Science Chimiche, Universit\`a di Padova, via F. Marzolo 1, I-35131, Padova (Italy)}

\author{Marta Rosa}
\email{marta.rosa@unipd.it}
\affiliation{ 
Dipartimento di Science Chimiche, Universit\`a di Padova, via F. Marzolo 1, I-35131, Padova (Italy)}  
\affiliation{Center for Biomolecular Nanotechnologie, Istituto Italiano di Tecnologia, Via Barsanti, I-73010 Arnesano, LE, Italy}

\author{Stefano Corni}
 \email{stefano.corni@unipd.it}
\affiliation{ 
Dipartimento di Science Chimiche, Universit\`a di Padova, via F. Marzolo 1, I-35131, Padova (Italy)}
\affiliation{Padua Quantum Technologies Research Center, Universit\`a di Padova}
\affiliation{
Istituto Nanoscienze—CNR, via Campi 213/A, 41125 Modena (Italy)
}

\begin{abstract} 
We develop an hybrid quantum-classical algorithm to solve an optimal population transfer problem for a molecule subject to a laser pulse. The evolution of the molecular wavefunction under the laser pulse is simulated on a quantum computer, while the optimal pulse is iteratively shaped via a machine learning (evolutionary) algorithm. A method to encode on the quantum computer the n-electrons wavefunction is discussed, the circuits accomplishing its quantum simulation are derived and the scalability in terms of number of operations is discussed. Performance on noisy intermediate-scale quantum devices (IBM Q X2) is provided to assess the current technological gap. Furthermore the hybrid algorithm is tested on a quantum emulator to compare performance of the evolutionary algorithm with standard ones. Our results show that such algorithms are able to outperform the optimization with a downhill simplex method and provide performance comparable to more advanced (but quantum-computer unfriendly) algorithms such as Rabitz's \textcolor{black}{or gradient based optimization}.
\end{abstract}

\maketitle

\section{Introduction} 

The first proposals for a new paradigm of computation that would have gone beyond standard computer science date back to the works of Feynman and Deutsch \cite{feynman:1999, deutsch:1985}. In their seminal contributions for the first time it was argued that quantum mechanics could have been exploited, as much as classical physics, as a background for a new way of computing. 

Nowadays quantum computation is an extremely active field owing to the fact that such a paradigm of computation has been proven potentially disrupting and able to realize very challenging computational tasks in a faster and more efficient way than our classical most powerful computers \cite{arute:2019}. At the origin of this discipline there is the idea that a quantum system may be, by far, more suitable to simulate the properties of another quantum system \cite{feynman:1999}; it is equally true that the physical realization of quantum hardware can not be independent of the very precise knowledge of the underlying physics and of the capability to control its dynamics with an impressive spatial and temporal resolution \cite{debnath:2016, gambetta:2017}. From these considerations naturally comes to look at the development of new quantum technologies and the fundamental research in the field of atomic, molecular and optical physics \cite{castellanos:2020, blais:2020} as two mutually supporting processes.
An example of this philosophy is quantum optimal control theory (QOCT) \cite{rabitz:2000, brif:2010} where the knowledge of the system (its assumed Hamiltonian) is sufficient to suitably shape, both in time and space, an external perturbation to achieve a desired response of such system. Such a strategy has been applied in several fields ranging from photochemistry to quantum gate synthesis \cite{zhu:1998,herek:2002,mogens:2020}; notably, with regard to the last example, QOCT is generally conceived as a tool to improve quantum devices performance while we rarely think of a quantum computer (exception made for closed-loop feedback control experiments as in \cite{riste:2012}) as an active party of the optimization process. In this work we put ourselves in this latter perspective by developing an hybrid algorithm which features quantum and classical computation to solve an optimal population transfer problem \cite{zhu:1995}; the approach we propose aims to complement, in a quantum computing fashion, the experiment proposed in the seminal work of Judson and Rabitz \cite{judson1992teaching}. They demonstrated an optimal control protocol based on the experimental measurement of a property disclosing the molecular wavefunction evolution in presence of an external light pulse, coupled to a learning procedure that shapes the control pulse to obtain the desired excitation. In the present work the quantum computer replaces the experimental apparatus, extending the range of application of this approach to those systems for which setting up such an experiment from scratch is not feasible (e.g., molecules undergoing photodamages during the experimental optimization, molecules not synthesized yet). The choice of this particular case has been also motivated by the increasing interest in the ultrafast dynamics of electronic states that have been probed experimentally with ultrashort pulses in single molecule experiments \cite{accanto:2017}. \textcolor{black}{Moreover, here, we intend to provide a further example of the application of quantum computation methods to chemistry. A great effort, to date, is directed to the development of algorithms for the characterization of properties concerning the electronic structure of molecular systems \cite{Cao2019}; although their importance is primary, we hope that this work can demonstrate how new quantum technologies can be exploited to address time-dependent problems to complex systems by combining state-of-the-art quantum chemistry methods with quantum simulation techniques.} 

\begin{figure}

\definecolor{arancio}{HTML}{E89005}
\definecolor{azzurro}{HTML}{457B9D}
\definecolor{azzurrino}{HTML}{A8DADC}
\definecolor{tartaruga}{HTML}{B1E03C}

    \centering
\usetikzlibrary{arrows.meta}
\tikzset{%
  >={Latex[width=2mm,length=2mm]},
            base/.style = {rectangle, rounded corners, draw=black,
                           minimum width=8cm, minimum height=4cm,
                           font=\sffamily, fill=blue!30!green},
  activityStarts/.style = {base, minimum width = 4cm, minimum height= 0.75cm, fill=blue!30},
       startstop/.style = {base, minimum width = 2.5, minimum height = 1.5, fill=red!30},
    activityRuns/.style = {base, minimum width = 10, minimum height = 10, fill=red!30},
         process/.style = {base, minimum width=1cm, minimum height=0.75cm, fill=orange!15,
                           font=\ttfamily},
}

\begin{tikzpicture}[node distance=cm,
    every node/.style={fill=white, font=\sffamily}]
  \node (start) at (1, 1.35)  [activityStarts, fill=arancio!80]    {$E_{\textbf{a}}^{guess}(t)$, $\hat{\bar{\mu}}$ and EEs};

  
  \node (end) at (1, -0.75) [activityStarts,fill=tartaruga] {$|\Psi_{target}\rangle$ and $\textbf{a}_{opt}$};
  
  \node (In) at (-1, 0.01)  [process, below left=0.01cm and -1cm of start, fill=arancio!80] {In};
   
  \node (Out) [process, above right= 0.01cm and -1cm of end,fill=tartaruga] {Out};
  
  \node (ActivityDestroyed) at (-6, 0) [base, fill=azzurro!70] {};
  
  \node[draw, fill=azzurrino] at (-7.75,1.4) {Quantum Routine};
  
  \node (Quantum Routine) at (-7.75, 0) [startstop, fill=azzurrino] { \Qcircuit @C=.5em @R=.5em { & \multigate{2}{P_{|GS\rangle}}
& \multigate{2}{U_{\textbf{a}(t,0)}} & \qw & \meter \\
& \ghost{P_{|GS\rangle}} & \ghost{U_{\textbf{a}(t,0)}} &  \qw & \meter\\
 & \ghost{P_{|GS\rangle}} & \ghost{U_{\textbf{a}(t,0)}} & \qw & \meter 
}};

  \node[draw, fill=azzurrino, minimum width = 2.75cm, minimum height = 1.6cm, align = center, rounded corners] at (-3.75,-0) { if J $< \delta$ \\[2.4em] if J $\geq \delta$};

  \node (Classical Routine) at (-3.75,0) [activityRuns, fill=azzurrino] {$\max\limits_{\textbf{a}} J[\textbf{a}]$};
  
  \node[draw, fill=azzurrino] at (-3.75,1.35) {Classical Routine};
  
  \draw[->](-1,1.35) -- (-1.95,1.35);
  
  \draw[->](-4.45, 0.75) -- (-5.55, 0.75);
  
  \draw[->](-5.60, -0.25) -- (-5.10, -0.25);
  
  \draw[->](-3.05, -0.8) -- (-1.05, -0.8);
  
  \end{tikzpicture}
  
  \caption{\label{schema} Schematic diagram for the hybrid algorithm. The input is an initial guess field specified by the set of control parameters \textbf{a}, dipole moment matrix $\hat{\bar{\mu}}$ and excitation energies (EEs) of the molecular system of interest. The output is given by the set of control parameter $\textbf{a}_{opt}$, i.e. the optimal field. Parameters \textbf{a} in the initial ansatz directly enter in the quantum gate executing the simulation of the molecular dynamics $U_{\textbf{a}}(t,0)$. Subsequently they are adjusted
in a hybrid quantum-classical optimization loop until the functional $J$[\textbf{a}] is above a user-specified threshold. When this loop terminates, the resulting gate sequence ($P_{|GS\rangle}$, encoding of the initial state, $U_{\textbf{a}}(t,0)$, evolution) can also be used to prepare the state of the computer for further uses.}
\end{figure}
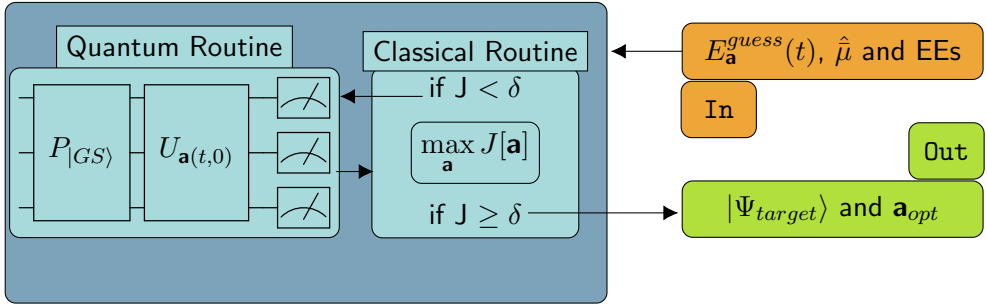

In the same spirit of Li \textit{et al.} \cite{li:2017}, where the authors focused on the problem of state preparation, we suggest the use of a quantum algorithm as a subroutine external to the classical optimization algorithm needed to evaluate the fitness function. Subsequently the classical optimization algorithm updates the set of control parameters feeding back again the quantum algorithm until convergence is reached (see Fig. \ref{schema}).  A similar approach has been also devised very recently by Magann \textit{et al.} \cite{Magann2020}. In particular, they present a thorough discussion of the numerical errors arising in the wavefunction evolution algorithm, estimate the resources needed with the proposed mapping and describe possible applications to vibrational and rotational control, and to the investigation of light-harvesting complexes. In this work we will focus on different aspects of the investigation with emphasis on the implementation of the routine on noisy devices, assessing the effect of quantum noise at different levels and motivating the choice of a genetic algorithm (GA) as an effective optimization routine.

The choice of the classical optimization algorithm is crucial for the performance of the whole procedure. Due to the quantum nature of the subroutine which computes the value of the fitness function after measure, the most natural choice is a gradient-free optimizer \textcolor{black}{(although a gradient based approach, extensively used in literature \cite{khaneja:2005}, will be considered along with the Rabitz algorithm as a reference for our calculations)}. As shown extensively in literature for the case of variational quantum eigensolver (VQE) \cite{peruzzo:2014,McClean2016} there are various choices for this kind of methods \cite{shen:2017, kandala:2017, colless:2018, moll2018quantum, Cao2019}; due to promising performance of machine learning in the field of multiparameters optimization, we chose to couple the quantum routine to a machine learning global search scheme, notably a genetic algorithm. Similar approaches have been already tested in the context of QOCT, but with a classical (as opposed to the present quantum) evaluation of the fitness function \textcolor{black}{\cite{chu:2001, zahedinejad:2014}}.

This paper is organized as follows: in Sec. \ref{theory_section} we present the theoretical tools needed to implement the hybrid algorithm; in particular discussing the main features of QOCT and the quantum simulation of a time-dependent molecular Hamiltonian with emphasis on the choice of the mapping adopted and on the quantum circuit synthesis. Subsequently, after having discussed the computational details in Sec. \ref{sec:comp}, Sec. \ref{numerical_results} is dedicated to analyze the numerical results assessing the performance of the mapping on noisy intermediate-scale quantum (NISQ) devices together with the implementation of the hybrid algorithm for the case of Cyanidin molecule; in particular, while providing a comparison between our hybrid solution and the result obtained with state-of-the-art algorithms on classical computers (gradient based optimization and the algorithm proposed by Zhu \textit{et al.} \cite{zhu:1995}), \textcolor{black}{we also analyze the performance of genetic algorithms when coupled to different noisy simulated quantum processors}. Finally, in the Conclusions we summarize the results obtained and consider possible future perspectives of this work.

\section{Theory} 
\label{theory_section}

\subsection{Proposed algorithm for quantum optimal control with quantum simulation of a time-dependent molecular Hamiltonian}

Quantum optimal control theory deals with the evolution of a dynamical quantum system undergoing an external manipulation, able to drive the system itself from a general starting state to a final excited state. The system evolution is described by the time dependent Schr\"odinger equation (TDSE), while the external perturbation which allows to control the dynamics is described adding a term to the isolated system Hamiltonian. In our case the extra term is a suitably shaped laser pulse, able to drive the system from an initial state $|\psi(0)\rangle = |\psi_0\rangle$ to a final state $|\psi(T)\rangle = |\psi_T\rangle$ corresponding to a chosen final time $t = T$. The final state should maximize the expectation value of a chosen operator $\hat O$ acting on the system. In this work, we will focus on a particular case, namely optimal population transfer: given a molecular system in the initial state we look for the shape of the laser pulse of length $t=T$ able to maximize the population of a specific target excited state. In other words, we want to tune the laser field profile in such a way that it provides the most precise, and experimentally attainable, optical excitation.
The optimal field can be obtained maximizing the following unconstrained functional:
\begin{equation}
  J[\textbf{a}] = \langle \psi_{\textbf{a}}(T) | \hat{O} | \psi_{\textbf{a}}(T) \rangle - \int_0 ^T \alpha(t) |E_{\textbf{a}}(t)|^2 dt
  \label{functional}
\end{equation}
Here $\hat{O} = |\Psi\rangle\langle\Psi|$ is the projector operator on the excited target state $|\Psi\rangle$ and the second term is a penalization factor that discourages the optimization algorithm to move towards high fluency fields; the time dependent factor $\alpha(t)$ allows to enforce a given envelope of the laser pulse. Such multiplier can be relevant e.g. when the optimal problem has to be solved for a particular experimental set up. Finally $|\psi_{\textbf{a}}(T)\rangle$ stands for the value of the state vector whose evolution is driven by a field $\bar{E}_{\textbf{a}}(t)$. The vector \textbf{a} specifies the set of control parameters.

There are two key-elements to be defined within QOCT algorithms: one is the approach to evolve the molecular wavefunction under the effect of the light pulse so to get $|\psi_{\textbf{a}}(T)\rangle$ (and thus $J$) for a given set of control parameters, the other is the optimization algorithm that iteratively leads to the best vector \textbf{a}. For the latter algorithm (the \textquotedblleft Classical Routine\textquotedblright in Fig. \ref{schema}), we adapted a machine learning scheme as described in the Method section.  Here we discuss the devised simulation of the evolution of the molecular wavefunction subject to the laser pulse, which is the  part running on the quantum computer (\textquotedblleft Quantum Routine\textquotedblright in Fig. \ref{schema}).  

Quantum simulation is a branch of quantum computation that aims to compute the evolution of a quantum system by programming the evolution of the state vector of a quantum computer such that they correspond to each other \cite{georgescu:2014}. In other words, let $U_{QC}(t,0)$ be the time evolution operator of the quantum computer (QC) and $U(t,0)$ the system time evolution operator we are interested into the computation of, we want to achieve:

\begin{equation}
    U(t,0) \quad {\xleftrightarrow[decoding]{encoding}} \quad U_{QC}(t,0)
\end{equation}

To this extent the problem of quantum simulation can be thought as a translation problem from the common language of quantum mechanics to the \textit{dialect} of quantum computation.

\paragraph{Encoding of the n-electrons wavefunction}
In this work we assume that a large enough set of electronic states of the molecule has been precalculated with some suitable quantum chemistry approach. Beside the energies of such states, the method provides also the matrix elements of the interaction operator with the external control. In the present case, since the control is the time-dependent electric field associated with a laser pulse, the required operator is the dipole moment operator $\bar{\mu}$.

In this section we describe the method that we have used to map a molecular wavefunction written in the (approximated) electronic eigenstate basis, e.g. configuration interaction (CI) \cite{helgaker:2014}, into the state of the quantum computer. 

First of all, for sake of clarity, we recall the common notation for an arbitrary state of the quantum computer composed by an M-qubits register given by:

\begin{equation}
    |\Phi\rangle = \sum_{\beta_1,\dots,\beta_M} c_{\beta_1,\dots,\beta_M} \big( \bigotimes_{i=1}^M |\beta_i\rangle \big)
\end{equation}
Where $\beta_i$ denotes the state of the single \textit{i}-th qubit which can be either $|0_i\rangle$ or $|1_i\rangle$ with the coefficients of the direct product states $c_{\beta_1,\dots,\beta_M}$ satisfying the normalization condition $\sum_{\beta_1,\dots,\beta_M} |c_{\beta_1,\dots,\beta_M}|^2 = 1$.

Our starting point is the molecular Hamiltonian \textcolor{black}{(i.e. the standard polyelectronic Hamiltonian containing electron-electron and electron-nuclei interactions)} in presence of an external field written as:

\begin{equation}
        \hat{H}(t) = \sum_k e_k |k\rangle \langle k| - \bar{E}(t) \boldsymbol{\cdot} \sum_{i,j} \bar{\mu}_{ij} |i\rangle \langle j|
        \label{hamiltonian}
\end{equation}
Where the indexes \textit{k},\textit{i},\textit{j} in the sums run over the possible n-electrons states approximated as the eigenvectors of the CI matrix; $e_k$ is the energy of the k-th electronic state, $\bar{E}(t)$ is a time-dependent electric field associated to the controlling laser pulse and $\bar{\mu}_{ij}$ is the transition dipole moment between the states $|i\rangle$ and $|j\rangle$ (or the corresponding permanent dipole when $i=j$).

With this approach we can write an arbitrary quantum state of the molecule as a coherent superposition of CI eigenvectors:

\begin{equation}
    |\Psi(t)\rangle = \sum_k^N c_k(t) |k\rangle
    \label{molecular_state}
\end{equation}

To find a link with the state of the quantum computer we encode the coefficient $c_k(t)$ (Eq. (\ref{molecular_state})) into the bitstring state coefficient of the quantum computer where all the qubits are in the state $|0\rangle$ but the k-th qubit which is in the state $|1\rangle$.

More formally we get the following equivalence:

\begin{equation}
        c_k(t) := \langle k | \Psi(t) \rangle \equiv \langle 0_1 \dots  0_{k-1} 1_k  0_{k+1} \dots 0_N | \Phi \rangle
    \label{mapping}
\end{equation}
Where we recall that $|\Phi\rangle$ is the state of the quantum computer and $|\Psi\rangle$ that of the molecule. Similar approaches have been recently considered for the simulation of bosonic Hamiltonians \cite{mcardle2019digital, ollitrault2020, sawaya2020resource}, here we extend this mapping to the simulation of many-fermions wavefunctions.
We may notice that this particular choice allows us to encode the coefficients corresponding to the different molecular electronic eigenstates into a particular subset of bitstring states where only one qubit at the time is in the state $|1\rangle$. This feature gives us the possibility to make a step further and establish a connection between the set of CI eigenvectors of Eq. (\ref{molecular_state}) and the Fock states of single occupancy for an N-fermions system. It is important to highlight that such an encoding exploits a \textit{formal} equivalence between a CI eigenvector and a single occupancy Fock state. Indeed on a physical point of view we know that we are dealing with very different objects: with the former representing an n-electrons wavefunction and the latter a single body quantum state. To conclude this parallelism we can think of such a mapping as an interpretation of the CI Hamiltonian (Eq. (\ref{hamiltonian})) as a single-body operator acting on the states of a quasi-particle, the different electronic states being the modes in which the latter can be found.

We can now exploit this equivalence to make use of the techniques that have been developed to perform the quantum simulation of a fermionic Hamiltonian \cite{somma:2002, whitfield:2011}. We chose to adapt the well known Jordan-Wigner (JW) mapping \cite{jordan:1928} to an n-electrons wavefunction by means of Eq. (\ref{mapping}). We recall that the JW transform allows to express fermionic operators in terms of Pauli matrices $\{ \sigma_x, \sigma_y, \sigma_z, \mathbbm{1} \}$ thanks to the following expressions:

\begin{equation}
\begin{split}
    a_j = -\bigotimes_{k=1}^{j-1} \sigma_k^{z} \sigma_{j}^{-} \quad \quad a_j^{\dagger} = -\sigma_{j}^{+} \bigotimes_{k=1}^{j-1} \sigma_k^{z}     
\end{split}
\label{JWmapping}
\end{equation}
Where $a_j$ and $a_j^\dagger$ are, respectively, the annihilation and creation operators for the \textit{j}-th particle while $\sigma_k^z$ and $\sigma_k^{\pm}$ are Pauli operators acting on the \textit{k}-th qubit.

Since we reduced the encoding of the n-electrons states into the single occupancy Fock space we can disregard the string of Pauli operators $\sigma^z$s leading to a much simpler expression. This simplification would not be possible if we had included in our encoding scheme the representation of a Slater determinant with a bitstring state with more than one qubit in the state $|1\rangle$, indeed, in that case, we should have taken into account permutation properties of the many body quantum states.

Finally, we end up with a simple equivalence between the k-th electronic state and the reference state of the quantum computer (i.e., when all qubits are in $|0\rangle$):

\begin{equation}
    |k\rangle \equiv \sigma_k^+ |0_1 \dots  0_{k-1} 0_k  0_{k+1} \dots 0_N\rangle = |0_1 \dots  0_{k-1} 1_k  0_{k+1} \dots 0_N\rangle
\end{equation}

By substitution of the previous expression in the molecular Hamiltonian (Eq. (\ref{hamiltonian})) we obtain:

\begin{equation}
    \hat{H}(t) = \sum_k \theta_{kk}(t) (\mathbbm{1}_k - \sigma_k^z)
    +\sum_{p>q}\frac{\alpha_{qp}(t)}{2}[\sigma_q^x\sigma_p^x + \sigma_q^y\sigma_p^y] = \Tilde{H}_z(t) + \Tilde{H}_{x,y}(t)
    \label{mapped_hamiltonian}
\end{equation}
Where, in the last equivalence, we put the subscripts to stress the dependence of the different terms on a particular subset of Pauli operators and adopted the following shorthands:
\begin{equation}
    \begin{array}{lr}
     & \theta_{kk}(t) = e_k - {\bar{E}}(t)\boldsymbol{\cdot}\bar{\mu}_{kk}  \\
     & \alpha_{qp}(t) = \bar{E}(t)\boldsymbol{\cdot}\bar{\mu}_{qp}
\end{array}
\end{equation}

\paragraph{Trotterization of a time-dependent Hamiltonian}
As we can see Eq. (\ref{mapped_hamiltonian}) is an explicit transposition of the time-dependent Hamiltonian in the $N$-electrons basis in terms of one-qubit operators (diagonal elements) and two-qubit operators (off-diagonal elements). We consider now how to derive the circuit representation of the time evolution operator $U_{QC}(t,0)$ based of such expression. It is well known \cite{mukamel:1999} that for an Hamiltonian such as the one of Eq. (\ref{hamiltonian}) the time evolution operator is given by a time-ordered expansion according to the equation:

\begin{equation}
    U(t,0) = \mathcal{T}e^{-i\int_0^t \hat{H}(t')dt'}
    \label{dyson}
\end{equation}
Where $\mathcal{T}$ is the Dyson's time ordering operator.

To compute the time evolution operator by making use of the Trotter-Suzuki (TS) approximation \cite{lloyd:1996} we discretize the time axis in $K$ time slots of width $\Delta t = t_{j+1} - t_j$ and approximate the integration as if the Hamiltonian were constant within the given time interval:

\begin{equation}
    U(t,0) \approx \mathcal{T}e^{-i\sum_{j=0}^K \hat{H}(t_j) \Delta t} 
\end{equation}
We are now able to decompose the time evolution operator $U(t,0)$ as a stepwise time-independent evolution and to apply the TS decomposition to each single-step evolution operator:

\begin{equation}
      U(t,0) = \prod_{j=0}^K U(t_{j+1}, t_{j}) 
\end{equation}

\begin{equation}
     U(t_{j+1},t_{j}) \approx  e^{-i\Tilde{H}_z(t_j) \Delta t}e^{-i \Tilde{H}_{x,y}(t) \Delta t}
     \label{singlestep}
\end{equation}
In the last equation we made use of the first order TS approximation but an extension to higher order approximation is straightforward \cite{wiebe:2010}. Notice that in the following we shall refer to this integration scheme for a time dependent hamiltonian as Trotter-Suzuki piecewise constant approximation (TS-PCA).

The implementation of the first term of Eq. (\ref{singlestep}), i.e. the diagonal evolution of the molecular wavefunction, has been discussed in ref. \cite{whitfield:2011} consisting in the application of a phase gate on each qubit with the phase specified by $\theta_{kk}(t_j)$. Following the same reference the exponential of a string of Pauli operators has been implemented; in this work we considered three different implementations of the circuit accounting for $e^{-i \Tilde{H}_{x,y}(t) \Delta t}$ as summarized in Table I. The second circuit is exactly the implementation of the mapping we presented in the previous section where the string of $\sigma_z$ (CNOT cascade) has been disregarded. Furthermore we also considered a modification to the previous circuit in which all the gates accounting for the operations containing the exponential of $\sigma_x$s and $\sigma_y$s are performed separately in order to diminuish the single qubit operations needed to rotate the single qubit states into the correspondent $\sigma_x$/$\sigma_y$ eigenstate basis.

Finally, we highlight that on top of the TS approximation of Eq. (\ref{singlestep}) the off-diagonal evolution is also approximated by a first order product formula; in fact the Pauli string operators arising from different pair of states do not, in general, commute. As a consequence we have:

\begin{equation}
    e^{-i \Tilde{H}_{x,y}(t) \Delta t} \approx \prod_{\lambda} \Gamma(\alpha_\lambda(t_j)) = \prod_{\lambda} \Gamma_x (\alpha_\lambda(t_j)) \Gamma_y (\alpha_\lambda(t_j))
    \label{off-diagonal}
\end{equation}
Where $\Gamma$ is a shorthand for the boxed circuits (second row of Table I) and the last identity emphasizes the factorization of the circuit in two parts due to the $\sigma_x$ and $\sigma_y$ terms.

\begin{table*}[t]
\centering

\caption{\label{circuits} Quantum circuits representing the coupling of different electronic states in presence of an external perturbation. Here $q$ and $p$ correspond to different CI eigenstates and $\lambda$ is a shorthand for an arbitrary couple of states. In the first circuit a readaptation of the JW mapping in the n-electrons basis is shown; notice that with this approach an additional string of $\sigma_zs$ is present. Dashed lines of the CNOT gate stand for a CNOT cascade between all the qubits with label greater than $q$ and lower than $p$. Dashed boxes refer to the $\sigma_x$ contribution, while dotted boxes refer to the $\sigma_y$ part. Notice that the last circuit, unlike the previous ones, refers to the implementation of the whole operator $e^{-i \Tilde{H}_{x,y}(t) \Delta t}$.}

\begin{tikzpicture}[node distance=cm,
    every node/.style={fill=white, font=\sffamily}]
    
\node (table) at (0,0) {\begin{tabular}{m{60mm}m{80mm}}

Pauli representation & \quad Circuit \\
\hline\hline
\vspace*{0.9ex}
& \includegraphics[width=80mm]{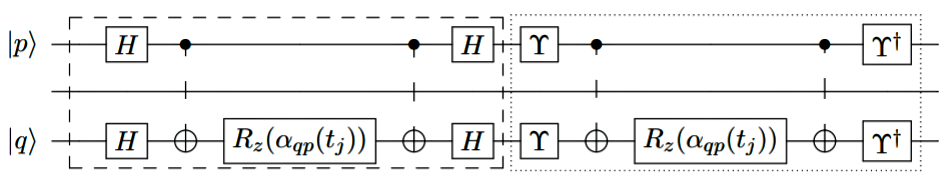}
\\ \\
& \includegraphics[width=80mm]{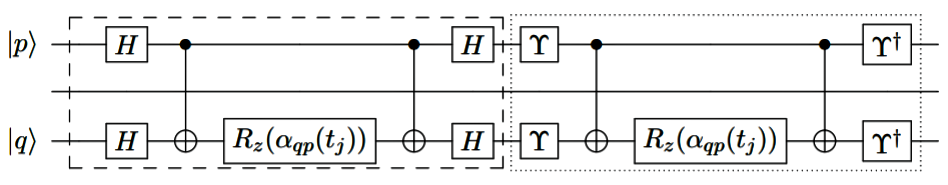} 
\\ \\
 & \includegraphics[width=80mm]{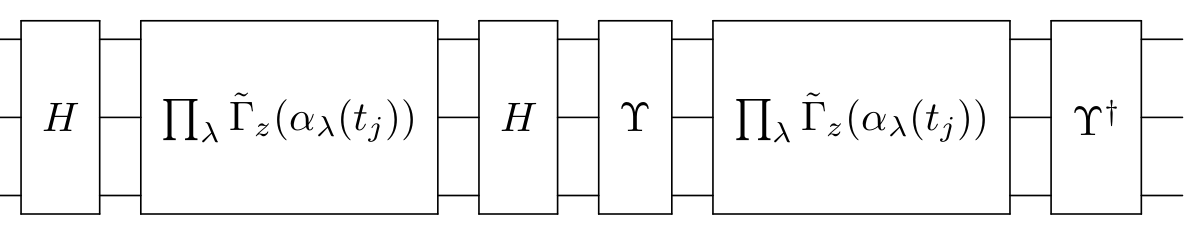}
\\
\hline
\end{tabular}};
    
\node (first_eq) at (-4.1,2.2) {$ e^{-\Delta t\frac{i}{2} \alpha_{qp}(t_j) [(\sigma_q^x\sigma_p^x +  \sigma_q^y\sigma_p^y) \bigotimes_{l = q+1}^{p-1} -\sigma_l^z]} $};

\node (second_eq) at (-4.97,-0.31) {$e^{-\Delta t\frac{i}{2} \alpha_{qp}(t_j) (\sigma_q^x\sigma_p^x +  \sigma_q^y\sigma_p^y)}$};

\node (terza_eq) at (-4.1,-2.85) {$\big [\prod_\lambda \Tilde{\Gamma}_x(\alpha_\lambda(t_j)) \big ] \big [\prod_\lambda \Tilde{\Gamma}_y(\alpha_\lambda(t_j)) \big ] $};
\end{tikzpicture}

\end{table*}

\paragraph{Quantum algorithm complexity}

A sufficient condition for an algorithm to be told efficient is that each step of its implementation must be accomplished in a number of operation and by requiring an amount of resources which increase polynomially with the system size \cite{sipser:2012}. Here we consider in more detail the complexity of our implementation to show that the number of operations scales polynomially with the dimension of the Hilbert space of the molecular $N$-electrons states. It is important to notice that, even though the implementation is efficient, the number of Slater determinants required still scales exponentially with the number of electrons and spin-orbitals used; as such we will also comment on other options that represent different compromises in terms of quantum hardware budget. We also point out the interesting analysis proposed in \cite{sawaya2020resource} where quantum circuits to switch encoding during the computation are presented and suggested as possible strategy to improve the ratio between efficiency of the implementation and quantum hardware requirements. Further we will not discuss neither the efficiency of the state preparation algorithm for a many-fermions state nor the efficiency of the evolution algorithm (Trotter-Suzuki decomposition) as they are both very well known techniques in the area of quantum simulation \cite{lloyd:1996, somma:2002}. To this extent, as a first point we consider the qubit efficiency of the encoding scheme; our mapping provides a linear scaling, with respect to the number of qubits, of the Slater determinants encoding. In fact, as we have pointed out previously, the electronic state $|k\rangle$ is mapped onto the bitstring state with all the qubits in state $|0\rangle$ but the k-th qubit. This implies that, considering the Hilbert space resulting from the tensor product of the individual qubits space, we are limiting our interest to a subspace made up of $Q$ bitstring states. Furthermore, even considering error correction procedures needed for fault-tolerant quantum computation, the qubit scaling remains linear since $\mathcal{O}$($Q$) additional qubits are needed for their implementation \cite{nielsen:2002}. Now, concerning the number of operations needed to simulate a single step evolution operator (Eq. (\ref{singlestep})), we will consider separately the contributions due to the diagonal elements of the Hamiltonian (i.e., $e^{-i\Tilde{H}_z(t_j) \Delta t}$) and the ones arising from the off-diagonal elements (i.e., $e^{-i\Tilde{H}_{x,y}(t_j) \Delta t}$). The former unitary operation needs exactly $Q$ single qubit operations to be performed; the latter, as shown in Table \ref{circuits}, has been considered in different versions. The first circuit shown in the table above, i.e. the one not considering any constraint on the occupancy of our fictitious Fock states, represents an upper bound for the estimation of gate complexity of this implementation. Due to its hermiticity, this operator has $Q(Q-1)$/2 independent terms, each of which can be implemented with $2Q$ two-qubits operations, as discussed in detail in ref. \cite{ovrum:2007}. Altogether we get a total gate count $\sim \mathcal{O}(Q^3)$. We can reduce the two-qubits operations complexity of one power if we consider a circuit in which we make explicitly use of the mapping boundary within the single occupancy Fock space (second row of Table \ref{circuits}). In this situation each term of Eq. (\ref{off-diagonal}) needs only 2 CNOT gates, leading to a total gate count of at most $\sim \mathcal{O}(Q^2)$. Of course the matrix associated to $\Tilde{H}_{x,y}(t_j)$ may be sparse, reducing the scaling or at least the prefactor. It may be worth noticing that these considerations do not take into account that the actual implementation of a two-qubits gate on the hardware could require additional $\mathcal{O}(Q)$ operations, as not all the qubits in the quantum processor interact directly between each other. On the downside, this enconding only exploits a subset of the computational space of the quantum hardware. Specifically, the dimension of such sub-space is the same as the number of qubits. Anyway, the idea can be extended by considering mapping to larger subspaces, such as double, triple, etc. excitations vectors (e.g., vectors with all the qubits 0 but two or three or more). The dimension of such space is the binomial coefficient ${Q}\choose{m}$, where $m$ is the excitation order. While using a larger sharing of computational space, such subsets require a larger number of CNOT gates to encode each term of Eq. (\ref{off-diagonal}), since in the $m$-excitations subspace one needs strings of $2m$ Pauli matrices (with $m \le Q/2$) to specify a particular bitstring state; each term is a sum of $2^{2m-1}$ of such strings. Depending on the features of the quantum hardware and the problem at hands, the best $m$-excitation subspace may be chosen.

Other options for the mapping can be considered as well, such as the use of a binary mapping to encode the Slater determinant integer label (sorted, for instance, according to increasing energy ordering) and the bitstring state of the computer. In this case we will have a qubit register dimension of $\log_2Q$. This option has been considered in \cite{Magann2020}; even though more efficient from an information storage standpoint, this mapping requires more complex expressions for the Pauli representation that, in turn, imply a greater number of two-qubit gates. Again, we notice that the choice of the mapping should be done carefully evaluating the balance between circuit depth, classical and quantum computational resource requirements.

To conclude this discussion we recall that, in the particular application considered in this work, i.e., the quantum simulation of a time dependent Hamiltonian, the total complexity must take into account also the number of propagation steps $K$ needed. As a consequence, the final estimation of the two-qubits operations for our circuit is $\sim \mathcal{O}(Q^2K)$, still ensuring the efficiency (as defined above) of the implementation.

\subsection{Optimization of the $J$ functional} 
\label{methods}
Different approaches are available to optimize the $J$ functional (Eq. (\ref{functional})) with different pros and cons depending on the system under study and the purpose of the optimization \cite{brif:2010}. A notable example is represented by the well established Rabitz algorithm \cite{zhu:1998}, which shows excellent convergence properties both in vacuum and in the presence of a solvent \cite{rosa:2019}. Despite its success, this algorithm is intrinsically unsuitable to be used coupled with a quantum simulation algorithm, as it requires the knowledge of the state of the system at each instant of its evolution. This means we should measure the state of the quantum computer at each time step, nullifying the advantages of the quantum procedure. Notwithstanding, in the choice of the final algorithm to be used for our procedure, we have kept Rabitz algorithm as a reference to asses the quality of our results.

In this work we follow the idea of Li \textit{et al.} \cite{li:2017} working
out an hybrid algorithm in which quantum computation accounts for propagating the system wave function under the influence of the external field, while an external classical algorithm optimizes a set of control parameters for the field. The reason behind our choice is that, doing so, we are able to assign to the quantum processor the computationally most demanding part of the algorithm (i.e. the propagation of the wavefunction and consequently the evaluation of the functional $J$) leaving to the classical computer the task of choosing how to proceed in the control parameter landscape exploration. Nevertheless, exploitation of quantum search algorithms in conjunction with the quantum simulation circuits discussed above (see Table \ref{circuits}), could allow a further speedup of this implementation. The last point is still an open problem and could be subject of a future work.

Differently from Li \textit{et al.} \cite{li:2017}, we focus on classical optimization algorithms that do not require the knowledge of the gradient of the objective function with respect to the optimization parameters. As a matter of fact, calculating the gradients adds extra noise to the procedure, as noted for VQE \cite{mcardle:2020}. A first possibility that we considered for the classical optimization procedure is the Nelder-Mead (NM) algorithm \cite{nelder:1965} which has been applied successfully in the CRAB and DCRAB methods \cite{caneva:2011,rach:2015} and has been the choice for various hybrid quantum classical approaches to determine the electronic structure \cite{Cao2019}. This algorithm is a direct search method which first constructs a simplex in the multidimensional parameter space, and then modifies the simplex at each iteration under a set of predefined rules moving its centroid towards a minimum (hence the alternative name of downhill simplex method). Even though this method has been proved extremely robust \cite{nocedal:2006}, at the same time is affected by a very slow convergence rate, already considering few tens of control parameters \cite{machnes:2018}, as we will show in the Results section.

From a broader perspective, optimal control is an instance of optimization problems where machine learning (ML) has shown high potential \cite{mogens:2020}: for this reason, in order to improve the performance of the optimization algorithm, we decided to  implement a GA \cite{schwefel:1995} which is a direct search method evaluating multiple copies of the system and combining the best results to create a better generation, hence the name \textquotedblleft genetic algorithm\textquotedblright. In our case the algorithm propagates the system under the influence of different fields, and then the fields parameters are modified learning from the best
results obtained (see Sec. \ref{theory_section}). \\ We may notice that both in NM and GA methods a simple and massive parallel implementation is possible, as we are dealing with algorithms which rely on several independent evaluations of the functional \textit{J}. 

In both these options, we need to find a set of control parameters \textbf{a} that explicitly define the shape of the external perturbation (the field) in order to optimize the evolution of our system and the value of the functional \textit{J}. To obtain such a parametric dependence we decided to write the laser pulse as:

\begin{equation}
    \bar{E}_{\textbf{a}}(t) = \sum_{\alpha} \bar{u}_{\alpha} \big[ a_{0,\alpha} + \sum_j^M a_{j,\alpha} \sin(\frac{j \pi t}{T}) \big]
    \label{parametric_field}
\end{equation}
Where $\bar{u}_{\alpha}$ is a unit vector with direction specified by the index $\alpha$ running over the three cartesian components.

Here the field takes the form of a sum over different harmonics with frequency $\omega_j = \frac{j\pi}{T}$, where $T$ is the time duration of the laser pulse. The subscript emphasizes the parametric dependence on the set of amplitudes \textbf{a} which are, ultimately, our set of control parameters. We chose to use pre-determined $\omega_j$ values corresponding to Fourier frequencies, which have the advantage of describing a field which starts and ends at zero (provided it has no constant component $a_0$) thus more easily obtained with an experimental apparatus. Moreover, if a specific frequency range has to be considered due to particular experimental limitations, this is easily accomplished by choosing the terms to include in the summation of Eq. (\ref{parametric_field}). 
It may be worth noticing that alternative optimization algorithms implemented the optimization of the frequencies as well as of the amplitudes, in order to allow the systems to reach the global optimal solution \cite{caneva:2011}. In our work the optimization of the amplitudes alone was enough to escape local minima and reach a satisfactory optimal solution, thanks to the implementation of the evolutionary algorithm. Nevertheless, it would be possible both for the NM and the GA optimizers to include the frequencies in the optimization procedure. In the specific case of the GA it would be a straightforward procedure that would be easily implemented at the only price of increasing the number of optimization parameters.

\subsubsection{Genetic algorithms to optimize $J$}
\label{genetic}

Due to the quantum nature of the subroutine which computes the value of the fitness function, we decided to rely on a gradient-free optimizer. One of the most popular and transferable choices for a reliable algorithm in a multidimensional space are genetic algorithms, considered to be part of the broad family of machine learning techniques. 

The term genetic algorithms is due to the inspiration that these optimization strategies draw from the process of development of new species that takes place in nature \cite{schwefel:1995}. As the physical performance and the consequent success of an individual within a herd are related to the chromosomes, so the value of a function $f$(\textbf{x}) depends on the variables \textbf{x}. In this picture the performance of an individual (i.e. the value assumed by the function $f$) is completely specified by its genes (i.e. the value assumed by the set of variables \textbf{x}). In GAs several initial sets of parameters are generated (either by an educated guess or at random) thus forming a population of individuals; the optimization of the function is provided by letting the population evolve, forming a new generation of individuals, in accordance with the foundational rules of the evolution process (note that in our implementation the size of the population is kept constant, nonetheless alternatives with a varying number of individuals are possible). In our case each individual is a vector of amplitudes \textbf{a} describing the field used to evolve the wave function of the system in its final state. Generally speaking, the final state is the objective function of the algorithm and can be determined by numerical calculation or experimentally. Each of the individuals is then rated according to its fitness, which describes the success of the individual parameters. From our standpoint it is easy to understand that the fitness function corresponds to the $J$ functional. The fitness could be determined also from an experimental procedure, measuring any observable effect consequent to the electronic excitation, as charge transfer or light emission.\\ Concerning the mimicking of the evolution process, three main traits are usually considered in this framework: selection, recombination and mutation. 

\paragraph{Selection}
An example of biological merit is the transfer of one's own genetic kit: this is most commonly achieved by successful individuals. This behavior can be transferred in the implementation of a GA in different manners \cite{goldberg:1989, zeidler:2001}. We chose to build the next generation on top of the best $m$ individuals. It is important to notice that the ratio between the selected individual and the size of the population ($\frac{m}{n})$ is a critical parameter: indeed, if a too small number of individuals is selected, important features allowing to drive the optimization to the global minimum may be lost determining a premature convergence. Conversely, if too many individuals influence the next generation the successful traits could not emerge. This behavior would result in an optimization resembling a random exploration of the parameter landscape.

\paragraph{Recombination}

Once a subset of the population is selected to give birth to the new generation, the recombination process occurs. Here two individuals are chosen randomly between the selected ones and their amplitudes are mixed to give a new child vector. There are different flavours of recombination which allow more or less mixing of the two parent individuals. In our algorithm, each amplitude in the final new vector has 0.5 probability to come from one or the other parent. As a consequence, recombining the same two individuals a second time does not give the same child.

\paragraph{Mutation} 

With the last procedure (i.e., recombination) the rules to compute the offsprings are set. Nevertheless, eventually, a mutation can arise changing randomly a gene in a certain individual. In our implementation this phenomenon is performed summing a random number to an individual amplitude. Each random number is extracted from a gaussian distribution specified by a value of mean $\mu$ and standard deviation $\sigma$. We point out that the mutation is a non-deterministic process happening for each amplitude of each individual with a given probability $P$. In this implementation the choice of the three parameters (namely, $\mu$, $\sigma$ and $P$) is crucial to guarantee a sufficient exploration of the parameter landscape; in particular higher values of these three parameters lead to an higher chance and entity of the exploration.

Mutation and recombination can be combined in different ways to finally obtain the full population starting from the subset of selected individuals. Parameters for mutation and recombination can change during the evolution depending on the fitness behaviour. Generally speaking, recombination favours convergence toward a (local) solution while mutation, as already mentioned, favours exploration.

\hfill

The three steps described so far constitute the main blocks of the genetic algorithm we adopted in this work. It is important to notice that a genetic algorithm reaches the global minimum of a multivariate function, as at each iteration it acquires more information about the position of the minimum. The learning loop is actually made by the evolutionary process and it is efficient only when an accurate balance between recombination and mutation is achieved, avoiding, in this way, to get stuck in local minima or, on the other hand, to lose information due to an excessive mutation rate of the individuals.

Despite the number of choices needed to set up an evolutionary procedure, most of the optimization problems are robust with respect a large interval of values for the parameters, with the difference between different choices translating mainly in a faster or slower convergence towards the optimal solution, \textcolor{black}{see Appendix B for an account on the GA implementation performance with respect to increasing system complexity}. 
Our final choice was very simple: we recombine the subset of individual to obtain the full population (e.g. 60 starting individuals, 10 selected and recombined randomly to return 60 new individuals) and then we mutate all of them (i.e. each amplitude of each individual is mutated with a given probability). Several different choices could have been adopted to improve the efficiency of the algorithm and implement an automatic behaviour; nevertheless our purpose was to find a simple algorithm able to work efficiently combined with a propagation performed on a quantum computers. More details on the choice of the parameters are given in the following section. As shown in Sec. \ref{numerical_results}, the results obtained were satisfactory. 

\section{Computational details} 
\label{sec:comp}

All the calculations were performed with OCPy \cite{OCPy}, an homemade standalone Python software which features all the possibilities we have considered so far, i.e. hybrid or classical propagation coupled to different optimization algorithms, like Rabitz, GA, NM and Broyden–Fletcher–Goldfarb–Shanno \textcolor{black}{(BFGS)} methods.

\paragraph{Quantum chemistry calculations}
Geometry optimization of the cyanidin molecule was obtained using the NWChem software \cite{valiev:2010} at density functional theory (B3LYP, cc-pVDZ) level of theory. The first 10 electronic excited states were computed with GAMESS \cite{gordon:2005} at the CI single (CIS) level of theory using a 6-31G$^{**}$ basis set. The suitability of the number of excited states considered to evaluate the optimal dynamics involving a sufficiently large set of electronic states is assumed on the basis of a previous related work \cite{rosa:2019}. It may be worth noticing that, when QOCT is applied to an actual experimental situation , a more accurate description of the system is needed; the CIS method is usually affected by excitation energies overestimation \cite{dreuw:2005}, nevertheless an accurate description of the molecule electronic structure goes beyond the purpose of the present study.

\begin{figure}
    \centering
    \includegraphics[scale=0.2]{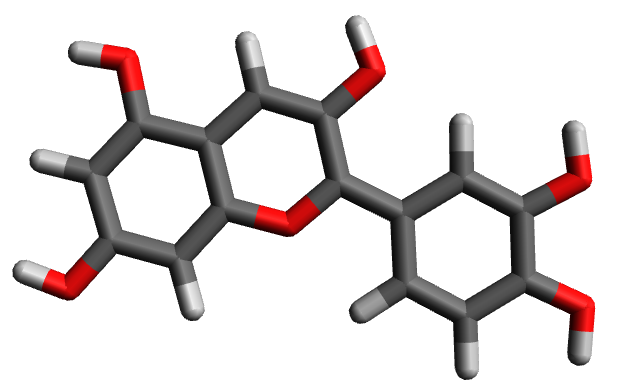}
    \caption{Optimized structure of the cyanidin molecule. Color code: red = O, black = C, grey = H.}
    \label{fig:my_label}
\end{figure}

\paragraph{Optimization procedure}

As previously mentioned, our aim is to maximize the population transfer between the ground state $|GS\rangle$, that we set as the initial state $|\psi_0\rangle$, and a given excited state, here the first excited state, $|1\rangle$. We applied the optimal control procedure both considering a manifold of eleven electronic states (the ground state plus the first ten excited states) and a smaller subset consisting in the ground state and the first two excited states. For sake of brevity, in the following we will refer to the former as system \textbf{Cyan11} and to the latter as system \textbf{Cyan3}.

The duration of the laser pulse is set equal to 250\,a.u., i.e. $\approx 6$\,fs. The reason to focus on this kind of perturbation is that a genuinely selective pulse is such if the electron dynamics is fast enough to occur in the manifold of electronic states unaffected by the consequent nuclear relaxation. In fact, as shown in Ref. \cite{klamroth:2006}, a $\pi$-pulse (i.e. a light pulse which causes population inversion in a two-level system) is not able to achieve the desired result in such a short time. In all cases in which we consider a classical method to compute the numerical propagation of the wavefunction we have adopted a Euler propagation method using a time step of 0.01\,a.u. (i.e. $\approx 2$\,$\times$\,$10^{-4}$\,fs); other, more robust, methods are possible (e.g., operator splitting technique in Ref. \cite{klamroth:2006}) nevertheless the choice of Euler propagator was done to keep the procedure as simple as possible and focus our discussion on the comparison between the classical and the hybrid implementation. \\
Concerning the parameters choice for the reference algorithms (\textcolor{black}{Rabitz and BFGS}), the initial guess field is a sinusoidal laser pulse resonant with the transition of interest ($|GS\rangle\longrightarrow|1\rangle$, $\omega = 0.125$\,a.u.) with amplitude set equal to 0.01\,a.u along all the three coordinates. To enforce small values of the field amplitudes we set the penalty factor $\alpha(t)$ equal to 10 (value in a.u., i.e. $e^2a_0^2\hbar^{-1}E_h^{-1}$) and kept constant throughout the evolution. NM calculations were performed with the initial guess simplex built upon the same guess pulse of the Rabitz algorithm by modifying per each vertex, one at the time, the value of a single component of \textbf{a}. More specifically, if the i-th component of the guess field is null it is automatically set to 0.00025\,a.u., otherwise it is increased by 5\%. 

In Table \ref{genetic_parameters} we show the genetic parameters adopted for our numerical calculations. We decided to alternate optimizations runs favouring exploration of the landscape and convergence runs towards the optimal solution. The only difference between the two is the value of $\sigma$ in the mutation procedure. A larger sigma means that the random numbers added during mutation span a larger interval, favouring exploration. In the exploration runs $\sigma$ value is equal to 0.001\,a.u., while in the convergence one it is equal to 0.0001\,a.u.. 
Our first and simpler choice was to perform only two runs, the first favouring exploration and the second favouring convergence. Depending on the system, a different number of iteration is needed: regarding \textbf{Cyan11} the total number of generations was 100 (50 generations of exploration and 50 generations of convergence), with a population size of 70 individuals. On the other hand, considering the case of \textbf{Cyan3}, the complete computation consisted of 30 generations (15 generations of exploration and 15 generations of convergence), with a population size of 40 individuals.
Subsequently, we worked in reducing the number of iterations and replicas in order to lower the computational effort when the propagation is performed on the 11 qubits computer.  Here we performed six runs made of 10 iterations alternating convergence and exploration until the algorithm achieves a threshold value for the target state population (that we set equal to 0.95). Then we end out optimization with a convergence run made of 40 iterations. Altogether the total number of iterations is still 100. Here there is a further difference between the exploration and convergence run, apart for the value of $\sigma$: indeed during the exploration phase a larger population size has been adopted (80 individuals), while the convergence run has been performed with a population size of 50 individuals. This tailoring of the optimization procedure was made specifically for the problem under study (i.e. system and hardware), and does not claim to be general or optimal. It was chosen on the base of the obtained performance and can be improved or replaced depending on the final purpose, usually exploiting a small number of test runs.
Finally, the value of the penalization factor for all the numerical calculations performed with the GA was kept constant and equal to 1\,a.u. as in this case to limit the intensity of the field is more effective to impose a maximum threshold for each amplitude equal to 0.005\,a.u..

\begin{table}[htbp]
\caption{\label{genetic_parameters} Genetic parameters adopted in the numerical applications. Different values have been considered for the complete set of excited state (first column) and for the smaller subset (second column). Likewise different parameters have been adopted to enhance a converging or exploring behavior. In brackets, when needed, values adopted in the hybrid implementation to reduce the computational cost.}
    \centering
    \begin{tabular}{lcc|cc}
    \toprule
   & \multicolumn{2}{c}{11 states} & \multicolumn{2}{c}{3 states} \\ \hline\hline
    & Exploration & Convergence & Exploration & Convergence \\ \hline
 Individuals   & 70 (80) & 70 (50) & 40 & 40 \\ \hline
 Sel. individuals   & 10 & 10 & 10 & 10 \\ \hline
 $P_{recombination}$  & 1 & 1 & 1 & 1 \\ \hline
  Mutated individuals & 10 & 10 & 10 & 10 \\\hline
  $P_{mutation}$ & 0.2 & 0.2 & 0.2 & 0.2 \\\hline
 $\mu$ & 0 & 0 & 0 & 0 \\\hline
 $\sigma$ & 0.001 & 0.0001 & 0.001 & 0.0001 \\
    \end{tabular}
\end{table}

\paragraph{Quantum propagation}

Quantum circuits accounting for the propagation were built thanks to the python library Qiskit \cite{qiskit}. The time step $\Delta t$ used for all our simulations is equal to 1\,a.u. (i.e. $\approx 2$\,$\times$\,$10^{-2}$\,fs), see Sec. \ref{numerical_results} for a more detailed account on the benchmark of the propagation. All the results concerning the hybrid implementation of the optimal control algorithm were obtained using the second circuit of Table \ref{circuits}. In all the computations performed on a quantum hardware we used the 5 qubits chip IBM Q X2 \cite{IBMQX2}; analogously all the noisy simulation were carried out including single qubit measurement errors, thermal relaxation and depolarizing noise tuned upon experimental parameters of the same quantum processor. Each circuit (both for quantum state tomography and quantum simulation) has been run 2048 times to build relevant statistics.

\textcolor{black}{Implementation of the hybrid algorithm in presence of a noisy simulation was accomplished considering four differents kinds of noise models. The standard approach to describe an error in quantum information science is to model the error event as a transformation of the qubit (or qubits when it is the case) state \cite{nielsen:2002}. These general transformations are usually named quantum channels. The bit-flip channel depicts a situation where for each operation we carry on the qubit register (i.e. one- or two-qubits gate) there is a probability $p$ for the state of the qubit(s) to be flipped, otherwise (i.e. with probability $1-p$) it is left unchanged. Similarly, the depolarizing channel refers to a transformation of the qubit(s) state to the maximally mixed state with probability $p$, else the identity applies.
These types of error have been combined to obtain the models summarized in Table \ref{Noise_models}.}

\begin{table}[htbp]
\caption{\label{Noise_models} \textcolor{black}{Noise models adopted in the numerical applications. Different error probabilities have been considered in order to assess the behavior of the genetic algorithm to the increase of the noise during the execution of the quantum simulation circuit. In brackets error probability for two-qubits operations. Depending on the different kind of noise used and their intensity we shall refer in the text to the different models with the following shorthands: bit-flip model (BF model); depolarizing error on single-qubit gates (SQ depol.(1) or SQ depol.(2), depending on the intensity); Mixed model when both the previous channels are included.}}
    \centering
    \begin{tabular}{lcccc}
    \toprule
    & BF model  & SQ depol. (1) & SQ depol. (2) & Mixed model \\ \hline
 Depolarizing channel probability   & 0 (0) & 1\,$\times$\,10$^{-5}$ (0) \qquad & 5\,$\times$\,10$^{-5}$ (0) \qquad & 5\,$\times$\,10$^{-5}$ (0) \\ \hline
 Bit-flip channel probability   & 1\,$\times$\,10$^{-5}$ (1\,$\times$\,10$^{-5}$) & 0 (1\,$\times$\,10$^{-5}$) & 0 (5\,$\times$\,10$^{-5}$) & 5\,$\times$\,10$^{-5}$ (5\,$\times$\,10$^{-5}$) \\ \hline \\
    \end{tabular}
\end{table}

\section{Numerical results}
\label{numerical_results}

In this section we provide the results of the numerical applications of the hybrid algorithm (and correspondent classical benchmark) to the molecule of cyanidin. Cyanidin is an interesting system where the excitation from the ground state to the first excited state $|1\rangle$ has been proved of fundamental importance for its role in the charge separation process in natural dye based solar cells \cite{ekanayake:2013}. \textcolor{black}{We applied the optimal control procedure to a cyanidin molecule, both considering a manifold of eleven electronic states (the ground state plus the first ten excited states) and a smaller subset consisting in the ground state and the first two excited states. For sake of brevity, in the following we will refer to the former as system \textbf{Cyan11} and to the latter as system \textbf{Cyan3}}. Showing the results for the system \textbf{Cyan3} will allow, not only to emphasize which dynamical features are lost when considering a too small subset of electronic states for the optimal problem, but also to establish a clearer connection with the results of the last section. In fact, to conclude the results section, we provide a performance analysis of the quantum simulation algorithm, referring to the particular case of system \textbf{Cyan3}, on IBM Q devices. 

\subsubsection{Benchmarking the quantum simulation circuit}

\begin{figure*}[htbp]
    \centering
    \includegraphics[width = \textwidth]{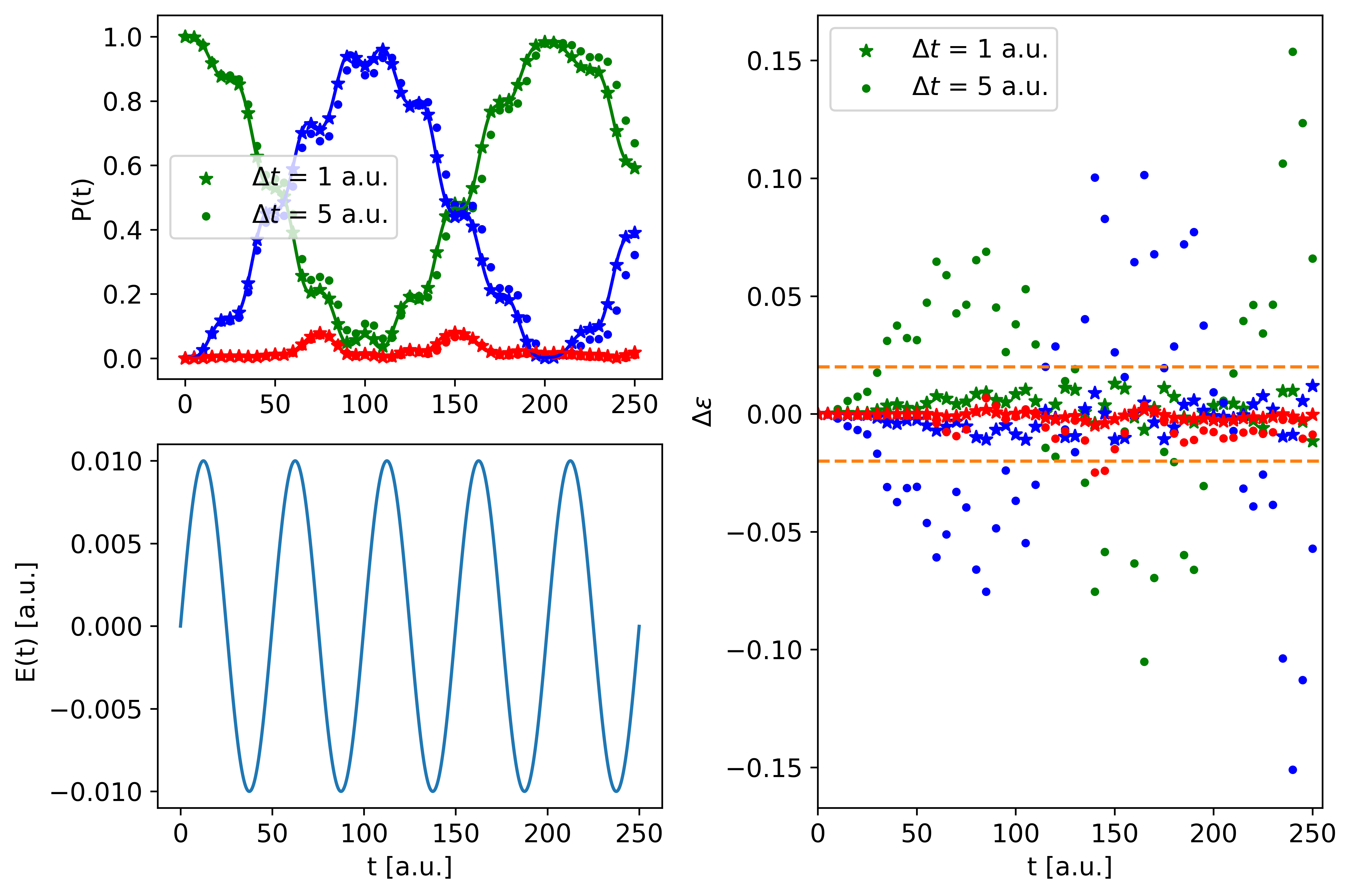}
    \caption{Upper panel: time evolution of the ground state (green) and first two excited states ($|1\rangle$, blue; $|2\rangle$, red) of cyanidin. Solid line represents the evolution computed with Eulero first order propagation ($\Delta t = 10^{-2}$\,a.u.). Stars ($\Delta t = 1$\,a.u.) and dots ($\Delta t = 5$\,a.u.) are computed simulating the quantum circuit representing the time evolution operator (See Table I). Bottom panel: laser pulse driving the evolution of the electronic states. Only one component is shown. (Right panel) Absolute deviation $\Delta \epsilon$ between the classical benchmark algorithm (Euler first order) and the simulated quantum circuits implementing the TS-PCA algorithm with different time steps. Dashed lines at $\Delta \epsilon = 0.02$ pose a treshold value for the minimum acceptable accuracy.}
    \label{circuit_benchmark}
\end{figure*}

In Fig. \ref{circuit_benchmark} we compare the time evolution (upper panel) of the first three states of cyanidin (for {\bf Cyan11}) under the influence of an ultrashort laser pulse (bottom panel) obtained with different methods. In particular we consider the dynamics computed integrating the TDSE with a classical algorithm (first order Euler), represented in figure by the solid line, and the evolution obtained adopting a simulated quantum circuit derived in the previous section based on the TS-PCA algorithm, see Sec. \ref{theory_section}. As a first consideration we may notice that the TS-PCA algorithm is able to reproduce the dynamics driven by the laser pulse; moreover it may be worth noticing the extreme robustness of the quantum algorithm with respect to the time step choice: our benchmark solution has been obtained with a time step $\Delta t = 0.01$\,a.u. ($\approx 2$\,$\times$\,$10^{-4}$\,fs), on the other hand, the evolution computed simulating the quantum circuit (stars in figure) represent the result obtained with a time step equal to 1 a.u. while the dots are obtained setting $\Delta t = 5 $\,a.u.. In both cases the simulated quantum algorithm recovers the qualitative information of the dynamics. To assess quantitatively the accuracy of the numerical solution we show in the right panel the absolute deviation $\Delta \epsilon$ of the TS-PCA propagation from the reference values of the classical algorithm using different time steps. A reasonable threshold for $\Delta \epsilon$ is 0.02 (dashed lines in figure) which is never exceeded when $\Delta t = 1 $\,a.u.; conversely we may notice that an higher time step, even though still useful for a qualitative inspection of the system, provides a much higher inaccuracy. For this reason all the results we will consider from now on have been obtained with a time step of 1\,a.u.. 

\subsubsection{QOCT: classical implementation}

\begin{figure*}[h!]

\begin{tikzpicture}[node distance=cm,
    every node/.style={fill=white, font=\sffamily}]
    
\node (figure) at (0,0)    {\centering
    \includegraphics[scale = 0.57]{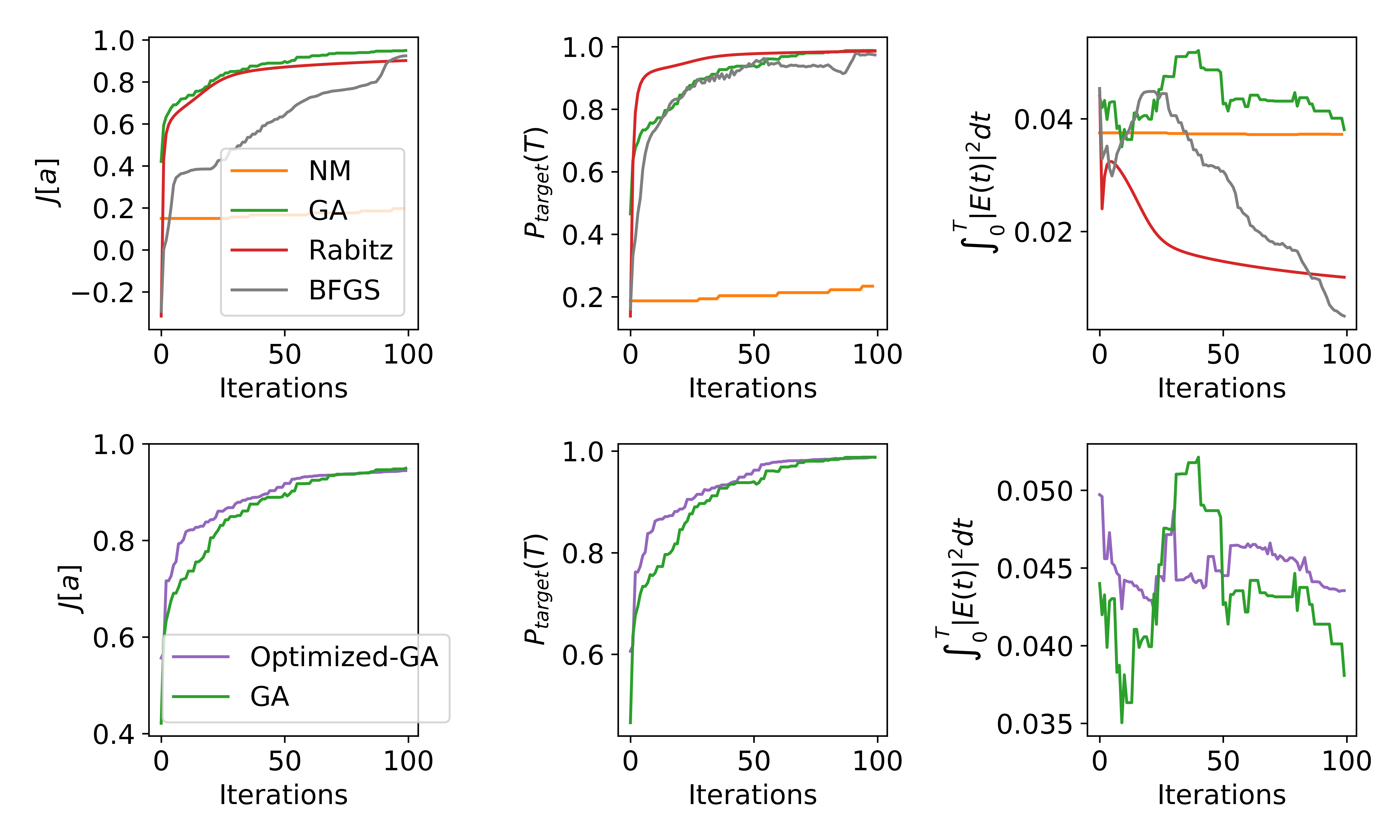}
};
    
\node (a) at (-7.3, 4.55) {a)};

\node (b) at (-1.75, 4.55) {b)};

\node (c) at (3.35, 4.55) {c)};

\node (d) at (-7.3, 0.1) {d)};

\node (e) at (-1.75, 0.1) {e)};

\node (f) at (3.35, 0.1) {f)};    
    
\end{tikzpicture}

\caption{\label{oc_comparison_classical} Optimal control applied to cyanidin with different approaches (fully classical implementation). Plots of $J$ (a,d), target state ($|1\rangle$) population (b,e) and field fluency (c,f). Bottom row plots show a comparison between two different strategies for the implementation of the GA; green solid line refers to the two steps strategy for balancing convergence and exploration, purple solid line refers to the less computationally demanding scheme which alternates multiple times greater or lesser exploration of the optimization landscape. All plots refer to system \textbf{Cyan11}.}

\end{figure*}

Fig. \ref{oc_comparison_classical} shows the results for the optimal problem applied to cyanidin molecule with a fully classical implementation. Here our aim is to compare the efficiency of the different approaches considered in this work. For this reason we will show only the results obtained for system \textbf{Cyan11}, as similar considerations apply for the simpler case of system \textbf{Cyan3}. Values for the functional $J$(see Eq. (\ref{functional})), target state population and field fluency are shown. As a first comment, we can see that the NM algorithm is completely outperformed by GA, BFGS and Rabitz algorithms; in fact, we can not appreciate a substantial improvement for any of the values reported. Actually, an additional order of magnitude of iterations are needed to achieve results of comparable quality to that of the other two algorithms. It is important to stress that this result is general, regardless of system specifics, and well established within the optimization community \cite{machnes:2018}. 

On the other hand the performance of the GA, even with our very simple strategy, is satisfactory and competitive with respect to the reference algorithms, as in all cases the optimization is capable of achieving an almost complete population transfer within the same number of iterations; regarding the value of the $J$ functional, it is important to notice that, as mentioned in Sec. \ref{methods}, we have adopted a different strategy for penalizing high fluency in the \textcolor{black}{reference algorithms}  and in the GA. \textcolor{black}{Concerning BFGS and Rabitz's algorithm}, we used $\alpha(t)= 10$\,a.u.), for the latter, we included no $\alpha(t)$ in $J$ and rather imposed an external maximum threshold for the amplitudes. Hence the discrepancy between Fig. \ref{oc_comparison_classical}a, where GA shows a higher value of $J$, and \ref{oc_comparison_classical}b and \ref{oc_comparison_classical}c, where Rabitz or BFGS results are equal or better than the ones obtained with the GA. Particularly on the field integrals values, it is clear that the best result is achieved by the BFGS algorithm. This last point should be interpreted considering that we are comparing a general purpose gradient-free algorithm (GA) with a gradient based optimization or with an algorithm whose fundamental working equations are problem dependent: such an accuracy could likely be achieved, within the framework of GAs, with a finer tuning of the parameters, which, again, is a problem dependent task or by increasing the number of iterations. An in depth analysis of the resulting electric field is presented in Appendix C.

Bottom row plots of Fig. \ref{oc_comparison_classical} show a comparison of the GA results obtained with two different strategies to balance convergence and exploration. The green solid lines in the upper and lower panel refer to the same results, obtained with the simpler two step exploration and convergence procedure; purple solid line shows the results for the same control problem but with the computationally less demanding scheme which alternates several times exploration and convergence (see Sec. \ref{methods} for details). This second choice will be adopted in our hybrid implementation to reduce the cost of simulating an eleven qubits device: as we can see they are almost equivalent both considering performance and final results. 

\subsubsection{QOCT: hybrid implementation using classical quantum emulator software}

\begin{figure*}[h!]

\begin{tikzpicture}[node distance=cm,
    every node/.style={fill=white, font=\sffamily}]
    
\node (figure) at (0,0)    {\centering
    \includegraphics[scale = 0.50]{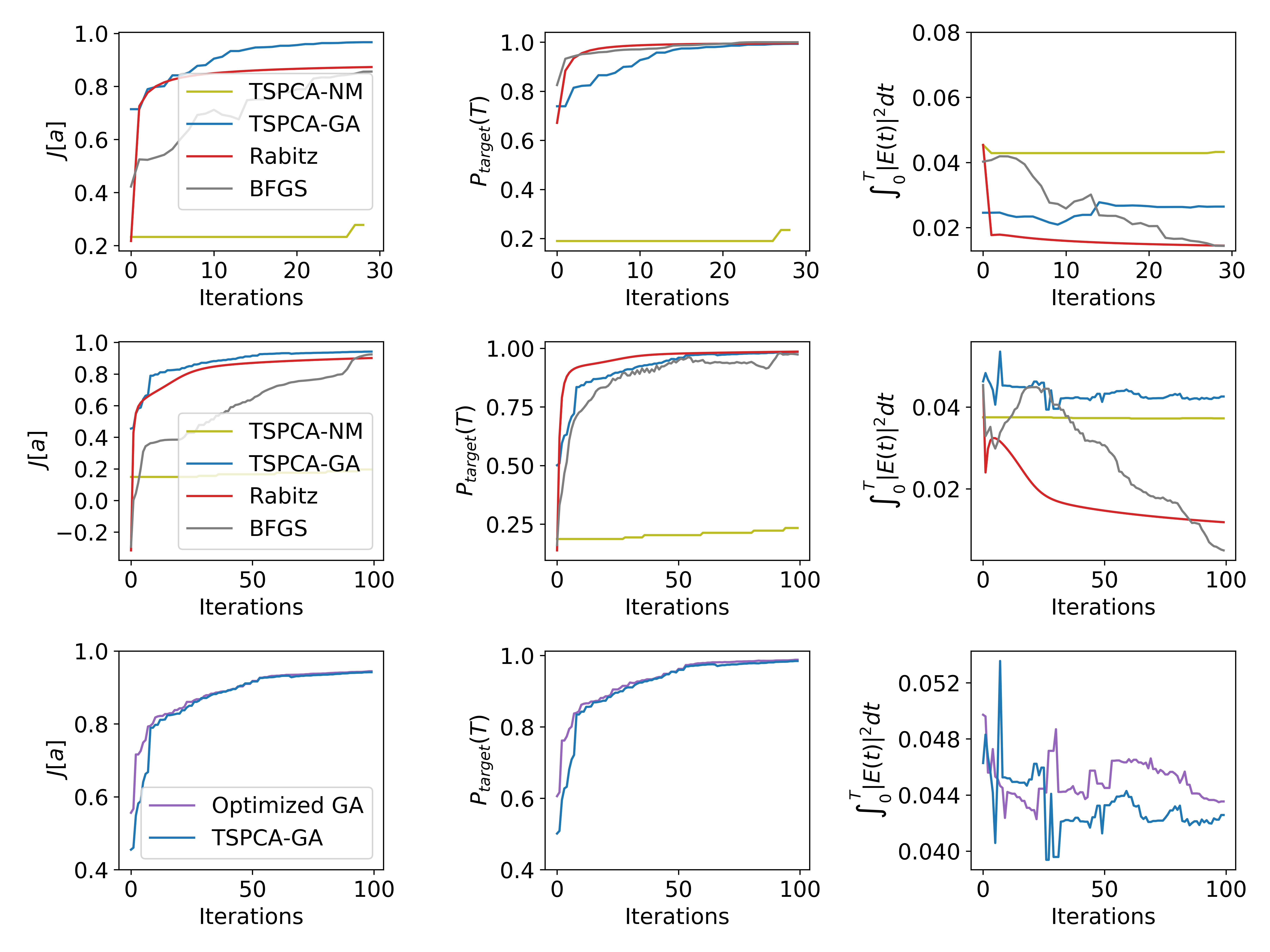}
};

\node (system 2) at (-8.70, 4) {\textbf{Cyan3}};

\node (system 1) at (-8.70, 0.35) {\textbf{Cyan11}};

\node (system 0) at (-8.70, -3.4) {\textbf{Cyan11}};
    
\node (a) at (-7.5, 6) {a)};

\node (b) at (-7.5, 2.25) {b)};

\node (c) at (-7.5, -1.70) {c)};

\node (d) at (-2.25, 6) {d)};

\node (e) at (-2.25, 2.25) {e)};

\node (f) at (-2.25, -1.70) {f)};

\node (g) at (3.05, 6) {g)};

\node (h) at (3.05, 2.25) {h)};

\node (i) at (3.05, -1.70) {i)};
    
\end{tikzpicture}

\caption{\label{oc_comparison} Optimal control applied to cyanidin with different approaches. Plots of $J$ (a-c), target state ($|1\rangle$) population (d-f) and field fluency (g-i). Color code: Purple - Classical implementation coupled to the GA optimizer; \textcolor{black}{Grey - Classical implementation coupled to the BFGS optimizer}; Blue - Hybrid implementation coupled to the GA optimizer; Yellow - Hybrid implementation coupled to the NM optimizer; Red - Rabitz algorithm. Upper row plots show the results for system \textbf{Cyan3}. Notice that all the results shown for the GA are obtained with the optimized scheme assessed in the previous section (See Fig. \ref{oc_comparison_classical}).}

\end{figure*}
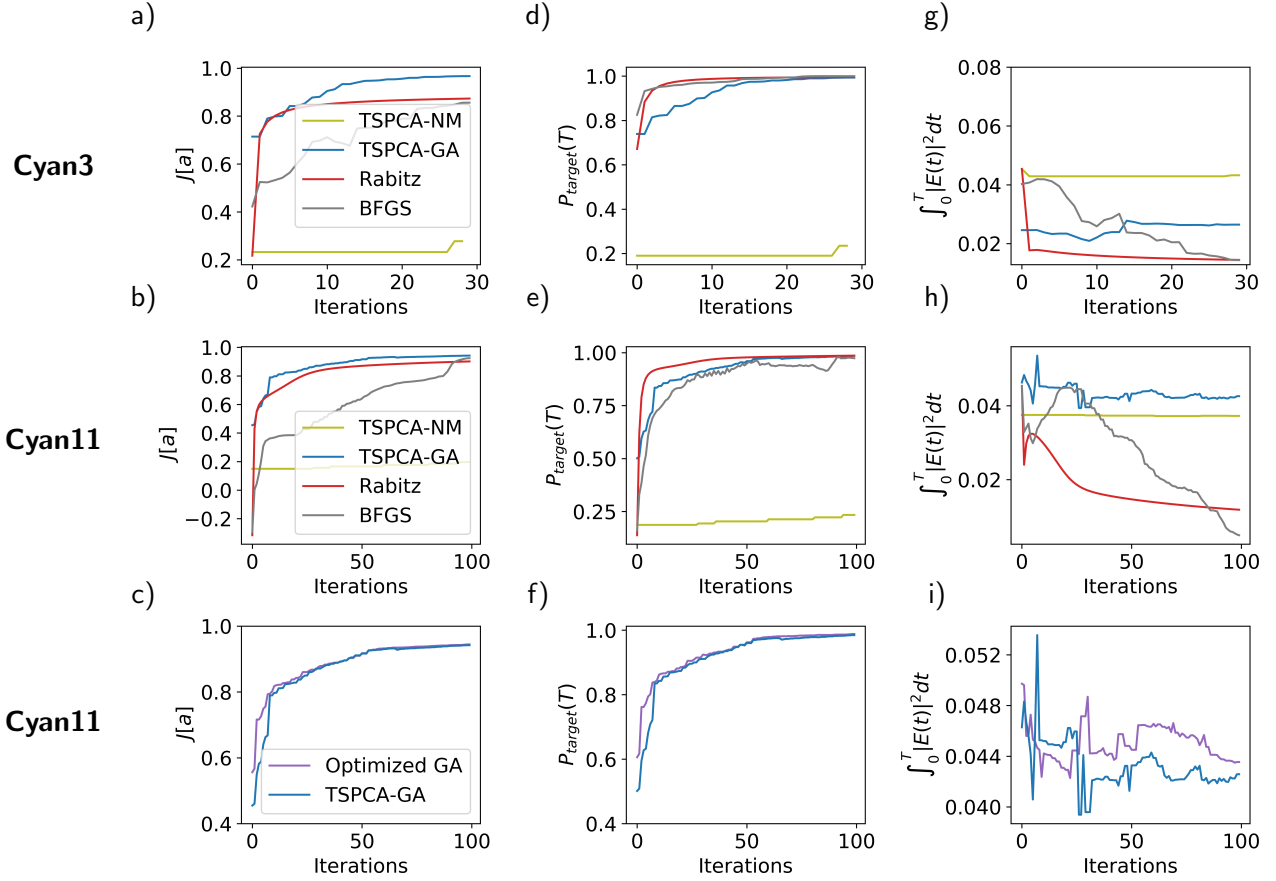

Aim of this section is to present the results of the hybrid implementations both for system \textbf{Cyan11} and system \textbf{Cyan3}. In particular Fig. \ref{oc_comparison} shows values for $J$, population of the target state and field fluency as a function of the iterations; upper row panels refer to system \textbf{Cyan3}, middle row panels to system \textbf{Cyan11} and finally bottom row panel show a comparison between the hybrid and classical implementation adopting GA as optimization strategy. Again we can see that also in the case of a simpler problem, i.e. optimal population transfer for system \textbf{Cyan3}, the NM algorithm displays a slow growth rate. Conversely \textcolor{black}{GA, BFGS and Rabitz algorithm} are able to reach a complete population transfer in fewer iterations. Regarding field fluency values of panel \ref{oc_comparison}g we can notice a smaller discrepancy (with respect to what shown in panel \ref{oc_comparison}h)  between GA and the reference algorithms that can be ascribed to the fewer control parameters (Eq. (\ref{parametric_field})) used than in the case of system \textbf{Cyan11}, see Appendix A for a discussion on the improvement of GA performances of field fluency values. It may be also worth noticing that in the case of this smaller system BFGS and Rabitz algorithm have an equivalent performance also regarding field fluency values. Concerning the results obtained for system \textbf{Cyan11} we can notice that all the considerations made in the previous section still hold. Finally, Fig. \ref{oc_comparison} bottom panel, we can appreciate the equivalence (in as much as we compare two results of a non-deterministic algorithm like the genetic one) between the classical and hybrid implementation as a further proof of the reliability of the quantum simulation algorithm. An analysis of the resulting optimized fields in terms of frequencies is given in Appendix C.

\subsubsection{QOCT: hybrid implementation simulating a noisy device}

\begin{figure*}[h!]

\begin{tikzpicture}[node distance=cm,
    every node/.style={fill=white, font=\sffamily}]
    
\node (figure) at (0,0)    {\centering
    \includegraphics[width = \textwidth]{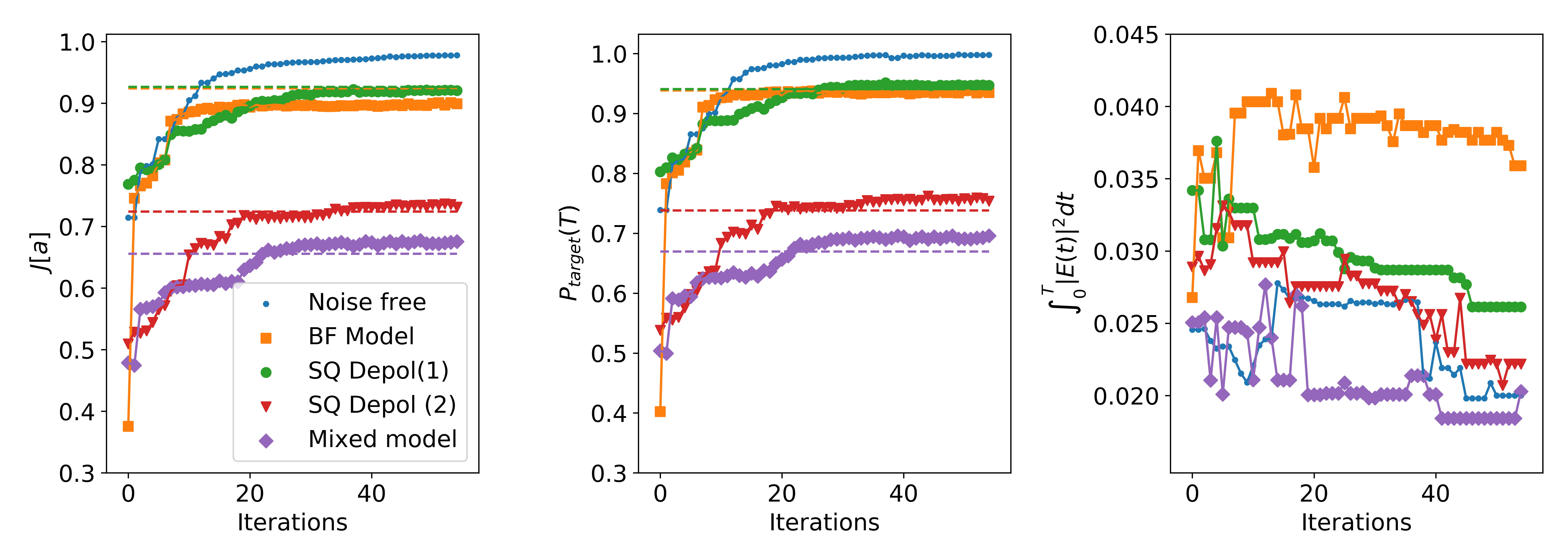}
};
    
\node (a) at (-7.95, 2.55) {a)};

\node (b) at (-2.4, 2.55) {b)};

\node (c) at (2.90, 2.55) {c)};

\end{tikzpicture}

\caption{\label{oc_noisy} \textcolor{black}{Optimal control with a noisy simulation of the quantum device. Plots of $J$ (a), target state ($|1\rangle$) population (b) and field fluency (c). Different noise models have been used to assess the robustness of the GA with respect to different error probabilities in the function evaluation. Dashed lines represent a reference value obtained with the optimal solution given by the BFGS optimization in presence of the corresponding noise model. All plots refer to system \textbf{Cyan3}.}}

\end{figure*}
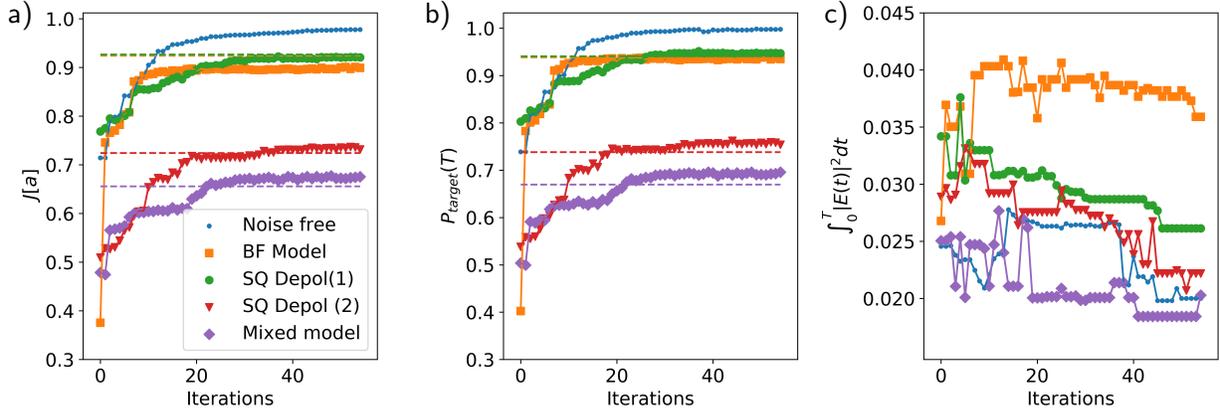

\textcolor{black}{Here we discuss the results regarding the solution of the control problem for \textbf{Cyan3} in presence of different noise models, see Fig. \ref{oc_noisy}. As mentioned in Sec. III, we considered four different noise models to assess the robustness of the GA with respect to a stochastic behavior of the function due to device imperfections. Single qubit error probability ranges from $1$\,$\times$\,$10^{-5}$ (BF Model) to $1$\,$\times$\,$10^{-4}$ (Mixed model) per gate execution, while two qubit operations reach a maximum error probability of $5$\,$\times$\,$10^{-5}$ (see Table \ref{Noise_models}). Even though our models assume much longer coherence times than the current ones (we have disregarded thermal relaxation errors), very recent contributions suggest that such models will be realistic for quantum processors available within a few years \cite{jurcevic:2020}. As we can see in Fig. \ref{oc_noisy} we reported values of the $J$ functional (\ref{oc_noisy}a), target state population (\ref{oc_noisy}b) and field fluency (\ref{oc_noisy}c). As the noise increases we can see how the optimization reaches convergence at lower values of the functional. Nevertheless it is likely that, depending on the function evaluation accuracy, the maximum value of $J$ reachable with the optimization decreases while not necessarily affecting the quality of the optimal result. To better understand this observation we reported (dashed lines in plots \ref{oc_noisy}a-b), for each different noise model, the value obtained by executing the circuit which simulates the evolution driven by the optimal pulse found with the BFGS algorithm in Fig. \ref{oc_comparison}. As we can see the GA approaches $J$ values very similar to the ones obtained in noise free conditions with a gradient based optimization. We may notice that in some cases (namely red and purple lines) the field obtained by performing the optimization in presence of noise is performing slightly better than the optimal field found by the BFGS optimization in noise free conditions; we suppose that this behavior can be related to the fact that the field found with a noisy optimization drives the quantum computer wavefunction through states more resilient towards depolarizing or bit flip errors. Further studies will be dedicated to address this point. On the other hand Table \ref{Optimal_noise_result_noisefree} provides the complementary data for this analysis: target state population at time $T$ obtained with the optimal field corresponding to different noisy optimizations. As we can see GA show excellent noise resiliency by providing fields that completely populate the target state. This feature could be provided by two concurrent factors: first, the algorithm is forced to retain the best solution found within a set of individuals (the functional grows regardless of the correctness of the propagation) and secondly each individual represents an independent propagation (i.e., more chances of obtaining physically meaningful information when the errors are not too big).}  

\begin{table}[h!]
\caption{\label{Optimal_noise_result_noisefree} \textcolor{black}{Target state population $P_{|1\rangle}(T)$ at time T for system  \textbf{Cyan3}. Evolution driven by the optimal field obtained as result of the control problem (see Fig. \ref{oc_noisy}) under noise free conditions.}}
    \centering
    \begin{tabular}{lcccc}
    \toprule
    & BF Model  & SQ Depol. (1) & SQ Depol. (2) & Mixed Model \\ \hline
 $P_{|1\rangle}(T)$   & 0.9980 & 0.9968 & 0.9977 & 0.9984 \\ \hline \\
    \end{tabular}
\end{table}

\subsubsection{Results on IBM Q hardware}

Up to now we presented results obtained on classical hardware, disregarding the effect of any noise in the quantum propagation \textcolor{black}{or considering mild effects that do not yet correspond to the current technological situation}. In this section we focus on the last issue. Due to the early stage of current quantum computer development, we were able to perform noisy simulations and actual computations on the IBM Q X2 hardware \cite{IBMQX2} only for the system \textbf{Cyan3}. We have limited our analysis to this simpler system as enlarging the electronic states subspace quickly prevents from distinguishing the characteristics of the different circuits implemented; in fact, due to the complexity of the investigated dynamics, the noise associated to the number of two qubits operations needed for a system larger than \textbf{Cyan3} would determine a complete suppression of the simulated wavefunction evolution. To quantify the effect of noise, we compute the fidelity between the density matrix reconstructed via quantum state tomography (QST) and a reference state obtained simulating the circuit in noise free conditions. The choice of the hardware (among the available IBM Q devices on cloud) has been dictated by the high inter-qubit connectivity required by the simulated hamiltonian; indeed, in order to minimize the number of additional gates needed to implement the circuits of Table \ref{circuits} (which assumes an all-to-all connectivity) in the actual computers, the device with the highest number of connections has to be preferred.

\begin{figure}[h!]

\centering

\begin{tikzpicture}[node distance=cm,
    every node/.style={fill=white, font=\sffamily}]
    
\node (figure) at (0,0)    {\includegraphics[scale = 0.625]{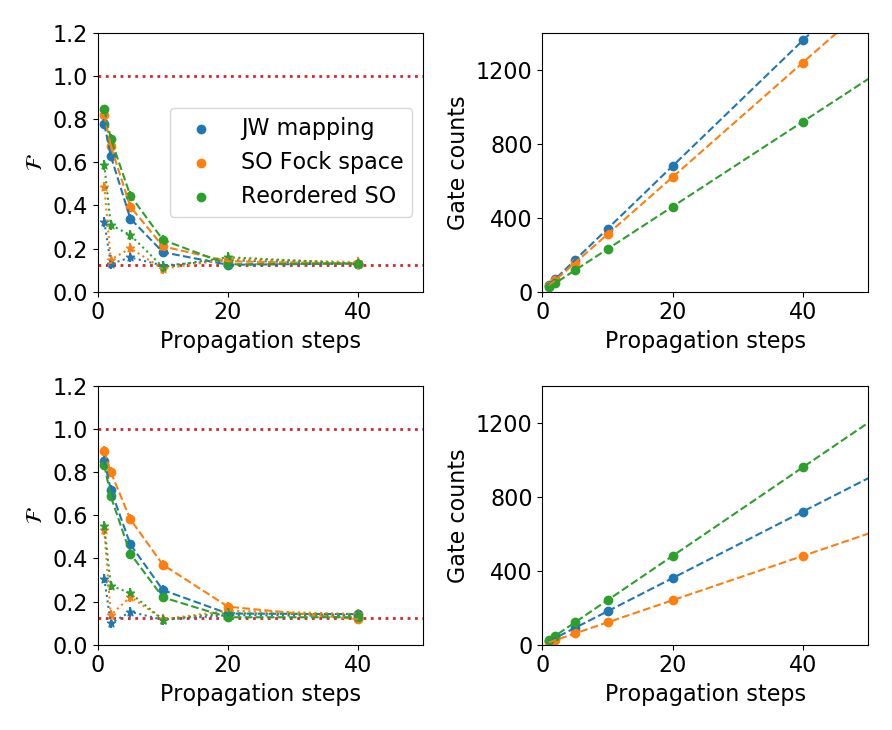}};
    
\node (a) at (-6.55, 5.25) {a)};

\node (b) at (1.15, 5.25) {b)};

\node (c) at (-6.55, 0) {c)};

\node (d) at (1.15, 0) {d)};

\end{tikzpicture}

\caption{(a,c) Fidelity of quantum state vs. Propagation steps for noisy simulations of system \textbf{Cyan3} (dashed lines) and actual experiments on IBM Q X2 (dotted lines); (a) Refers to the circuits in Table \ref{circuits}, (c) shows the results for the equivalent circuits after transpilation. Color code. Blue : Jordan-Wigner mapping readaptation in the n-electrons basis; Orange : single occupancy Fock space mapping; Green : Reordered single occupancy Fock space mapping. (b, d) Total gate count vs. propagation steps; same color code is used. Error bars represent the standard error of the quantum state tomography procedure with number of measurements N for each component of the density matrix N = 2048 assuming Gaussian statistics.}
    \label{fidelity_plot}

\end{figure}
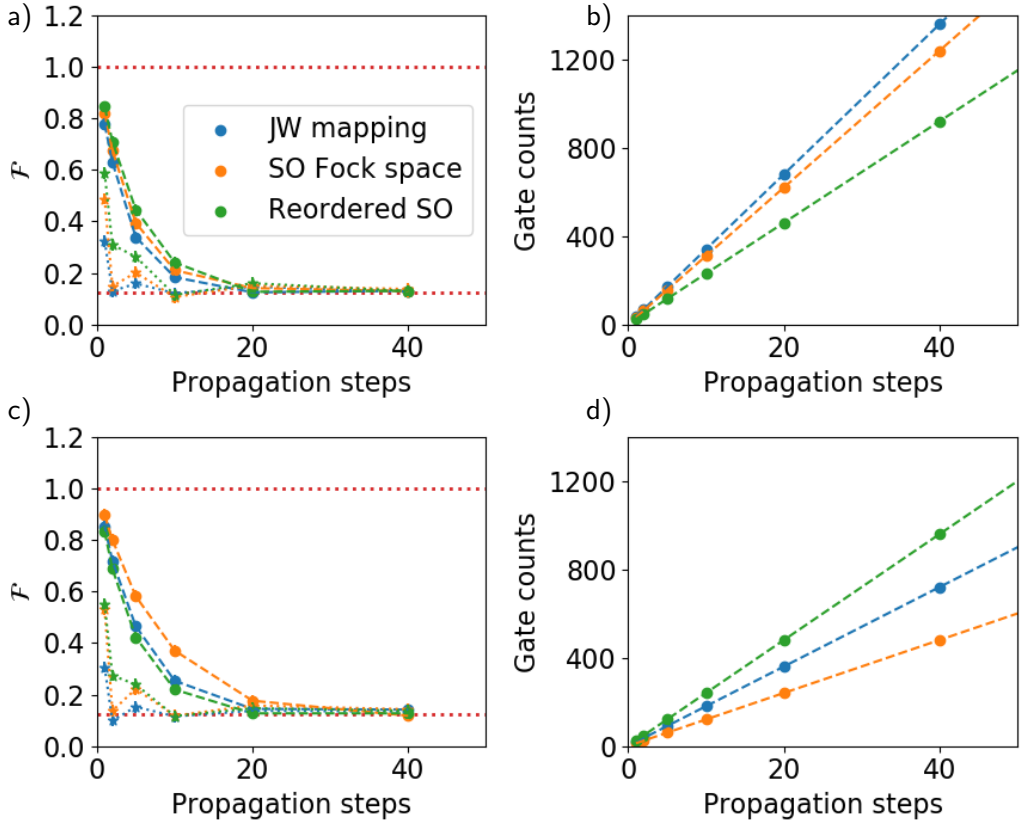

As a general consideration we observe, as expected, that as the number of gate counts increases (i.e. the depth of the circuit increases) we obtain a result farther from the classical benchmark; more explicitly, as mentioned in Sec. \ref{theory_section}, the number of operations grows linearly with the propagation steps (Fig. \ref{fidelity_plot}b-\ref{fidelity_plot}d) thus determining lower fidelity values (Fig. \ref{fidelity_plot}a-\ref{fidelity_plot}c). Our aim is to present these results as example to assess the technological gap that still has to be covered to exploit in this perspective our simulation algorithm. Clearly, as the depth of the circuit depends on the time duration of the pulse T, a profitable use of a quantum hardware with this algorithm is not likely at the present time; indeed, as we can see, after 20 steps of the propagation in all cases the state of the computer reaches a value of $\mathcal{F}$ equal to the value of thermal state that, according to Ref. \cite{zyczkowski:2005} is $\frac{1}{N}$ with N equal to the dimension of the Hilbert state of the quantum processor. Nevertheless these plots allow us to understand the importance of an accurate synthesis of the circuit as, exploiting the particular connection of our mapping with the single occupancy Fock space (second circuit in Table \ref{circuits}) we were able to obtain significant improvements. Indeed we can see that the readaptation of the Jordan-Wigner mapping on the electronic configuration space (upper circuit in Table \ref{circuits}) has a worse performance both in terms of fidelity and number of operations. To further improve the quality of the results we considered equivalent circuits obtained with an optimization routine (named transpilation) implemented in Qiskit which allows to reduce gate counts when a given circuit is mapped on to the topology of a particular hardware \cite{li:2020}; optimal compilation is a complex task that aims to minimize the performance loss due to the mapping on a limited connectivity hardware by exploiting gate cancellation and permutation rules to find the best virtual-to-physical qubit mapping and logical-to-native circuit mapping \cite{murali:2019}. As shown in the bottom panels of Fig. \ref{fidelity_plot} even though these improvements are significant when larger propagation steps are considered, we were not able to reach better fidelity values. We note that our reordered circuit (bottom circuit in Table \ref{circuits}) does not get any significant improvement by this procedure. As a further assessment, in Appendix D we discuss the probability distributions obtained measuring the computational basis states population, that confirm the considerations outlined here from the fidelity plots.

\section{Conclusions}

In this work we presented an hybrid approach featuring quantum and classical computation to solve optimal control problems in molecular systems. More thoroughly we adopted a quantum based routine for the evaluation of the cost function: to accomplish this task we proposed an efficient mapping of a multideterminantal wavefunction into the bitstring state of a quantum processor. A possible advantage of this construction, which is not based on physical principles, is that we can choose the criterion with which to assign the correspondence between the two sets of states thus exploiting hardware-specific information to improve the quality of the result. We will seek to explore these aspects in future works. As a case study we focused about the optimal population transfer between molecular electronic states of the cyanidin molecule. We compared the results of this problem both for different algorithms and for different choices of the classical optimization routine showing that machine learning approaches (here genetic algorithms), even without a fine tuning, can be competitive with respect to the well established and robust Rabitz's algorithm. The extension to optimization of other chemical relevant processes, such as photochemistry, may leverage on the on-going development of excited state nuclear dynamics performed on quantum hardware \cite{ollitrault2020}.

Moreover we discussed the practicalities of this implementation. From this standpoint we considered two different sets of states in which to follow the electronic dynamics, namely, a three-level system and a bigger set including up to the tenth excited state. The importance of discussing the results on a simpler system is manifold: the current fragility of quantum processors, even if equipped with a few dozen qubits, is put to the test not only by the implementation of very deep circuits, but also by maintaining coherence between qubits very far apart. Following the same idea in the last part of this work we have reported some tests with simple noise models and on IBM Q hardware to assess quantitatively the technological gap which has to be covered in order to exploit this quantum algorithm for many-electrons wavefunction evolution.

Efficient leverage of quantum properties of matter is an extremely challenging task calling for a common effort of the scientific community. By its part computer science, sustained by the increasing computer power, has provided chemists and physicists with the tools of machine learning which enable to extract information and find hidden patterns in data in an unprecedented way. With the advent of first quantum computers the development of quantum simulation algorithms must be explored to get the best out of both sides.

\begin{acknowledgments}
D.C. thanks Prof. Ida-Marie H{\o}yvik for useful discussions on cyanidin and the related quantum chemistry calculations. We acknowledge use of IBM Q Experience for this work. We also acknowledge funding from the ERC under the grant ERC-CoG-681285 TAME-Plasmons. 
\end{acknowledgments}

\setcounter{section}{0}
\setcounter{table}{0}
\def\thesection{\Alph{section}}
\def\thesubsection{A\arabic{section}.\arabic{subsection}}
\def\thesubsubsection{A\arabic{section}.\arabic{subsection}.\arabic{subsubsection}}
\def\thetable{A\arabic{table}}

\section*{APPENDIX A}
\begin{figure*}[h!]

\begin{tikzpicture}[node distance=cm,
    every node/.style={fill=white, font=\sffamily}]
    
\node (figure) at (0,0)    {\centering
    \includegraphics[width = \textwidth]{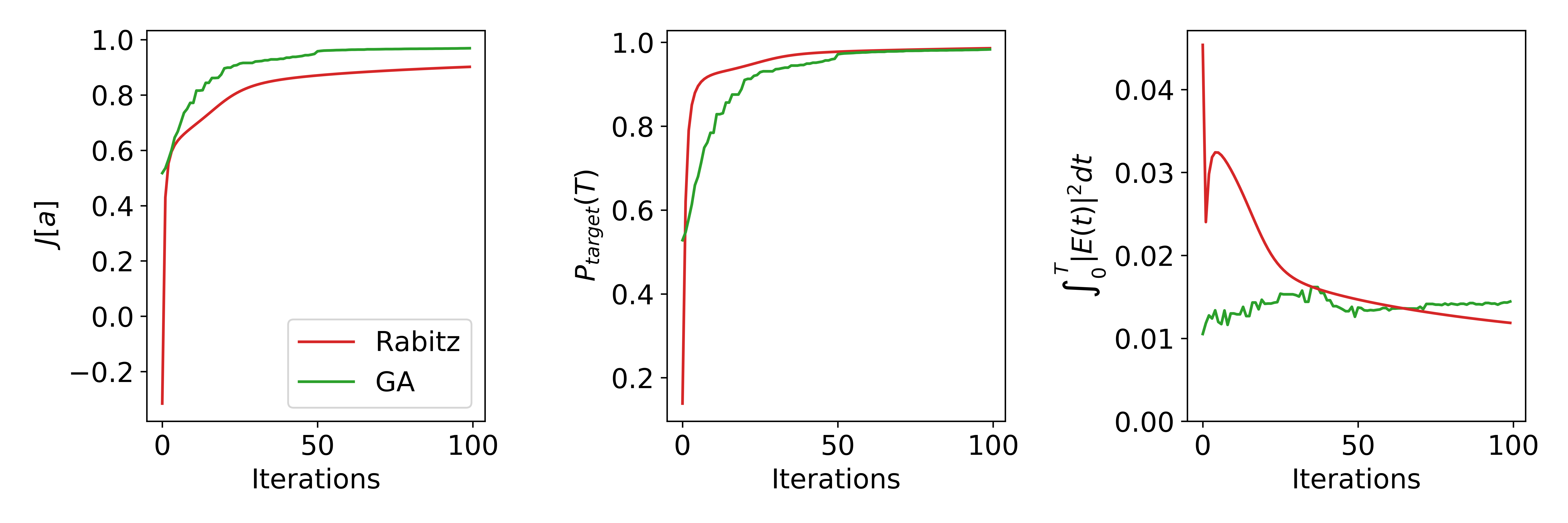}
};
    
\node (a) at (-7.85, 2.55) {a)};

\node (b) at (-2.4, 2.55) {b)};

\node (c) at (3.00, 2.55) {c)};

\end{tikzpicture}

\caption{\label{amplitude_limits}Optimal control applied to system \textbf{Cyan11}. Plots of (a) $J$, (b) target state ($|1\rangle$) population and (c) field fluency. Red solid line represent the reference result obtained with Rabitz's algorithm; green solid line shows the result obtained with the GA with a lower bound on amplitude values.}

\end{figure*}
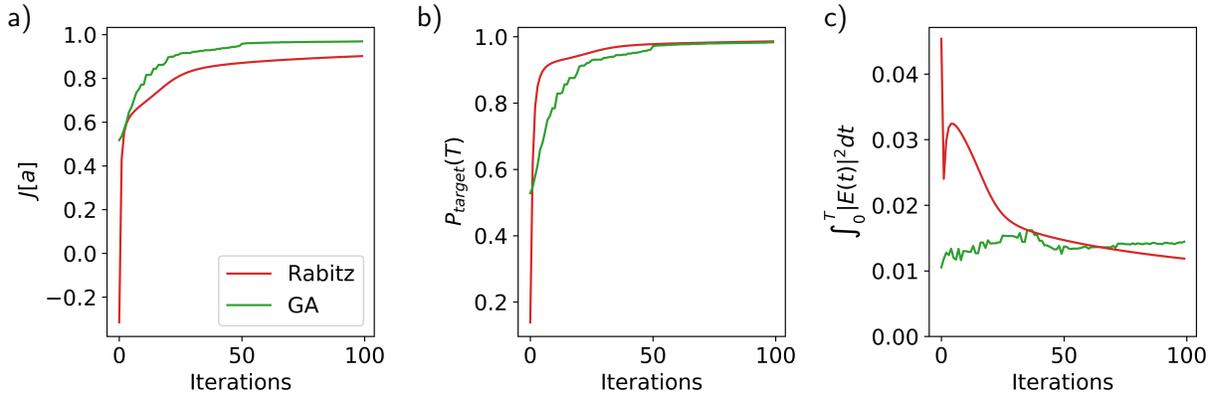

The optimization algorithms presented so far adopt different strategies to control the field fluency, which we want to keep as small as possible: Rabitz, BFGS and NM algorithm include in $J$ a scaling factor $\alpha$, which controls both the shape and the fluency of the field. The equation for $J$ is the same in the case of the genetic algorithm, nevertheless the algorithm itself enforce a limit to the value of the amplitude during their random generation and the mutation procedure. As a consequence, the scaling factor in $J$ can still control the shape of the field, but is quite useless in controlling the fluency. In Fig. 4 (main text) we decided to show results obtained with $\alpha$ = 1 for the genetic algorithm and $\alpha$ = 10 for Rabitz, with the purpose of not giving the false impression that the genetic algorithm was performing better than the Rabitz' one. In Fig. \ref{amplitude_limits} we show again the same comparison where the limit on the absolute value on the amplitudes in the genetic algorithm calculation is 0.0025 instead than 0.005. As it is possible to see, the result on the field is much more similar to the Rabitz one, without losing accuracy on the result on populations.

\section*{APPENDIX B}
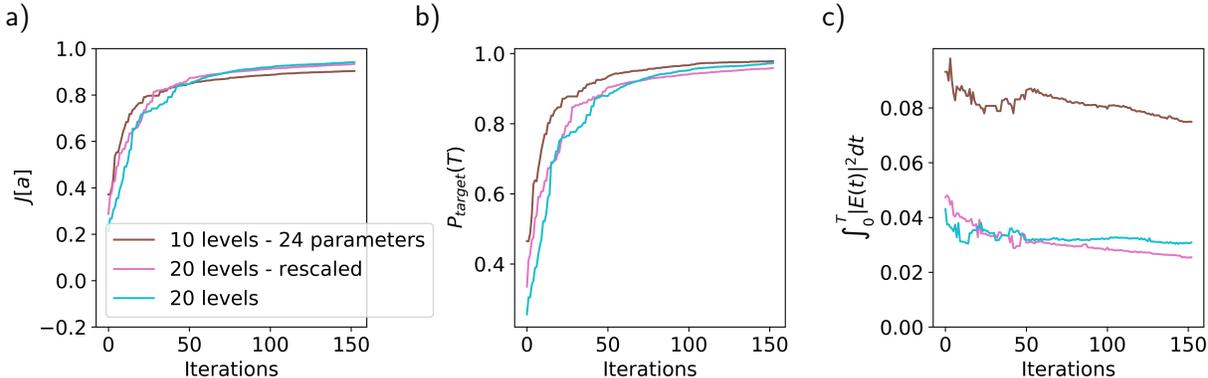
\begin{figure*}[h!]

\begin{tikzpicture}[node distance=cm,
    every node/.style={fill=white, font=\sffamily}]
    
\node (figure) at (0,0)    {\centering
    \includegraphics[width = \textwidth]{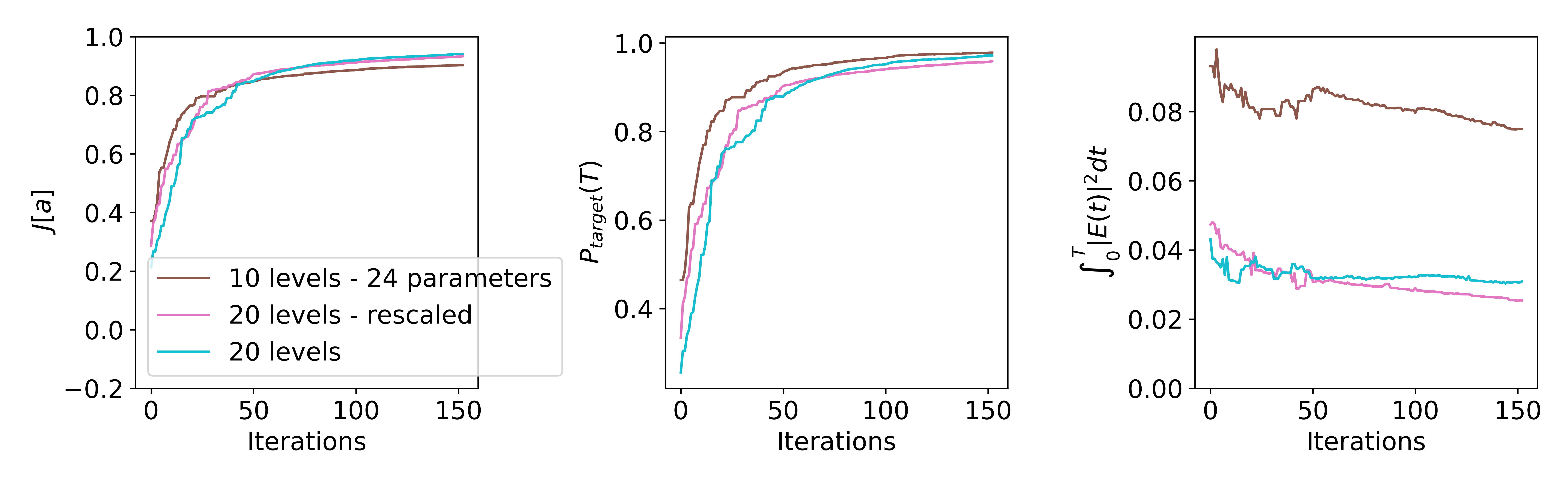}
};
    
\node (a) at (-7.85, 2.55) {a)};

\node (b) at (-2.4, 2.55) {b)};

\node (c) at (3.00, 2.55) {c)};

\end{tikzpicture}

\caption{\label{GA_complexity} Optimal control problem solution for the cyanidin molecule. Plots of (a) $J$, (b) target state ($|1\rangle$) population and (c) field fluency. Color code: cyanidin (20 levels) -  solid cyan line; cyanidin (20 levels with scaled energy spectrum) - solid pink line; cyanidin (10 leves with 24 optimization parameters) solid brown line.}

\end{figure*}

We performed some tests to understand how the very simple GA we implemented performs depending on the complexity of the system under analysis. Our interest was to assess which factors influence positively or negatively the results obtained.\\ 
To increase the complexity of the system, we decided to modify three different factors: the number of molecular excited states included in the calculation, the density of such states and the number of parameters to optimize with the genetic algorithm. Concerning this third point, the genetic algorithm is built to optimise the amplitudes of the field in Eq. (16), the number of amplitudes is three times the number of frequencies included in the summation, which in our calculation for cyanidin were chosen in an automatic way: the frequencies are such that the last one is higher than the last energy state. For the calculations performed in the main text, where we included 10 excited states, we have 12 frequencies plus the one corresponding to i = 0, which means 39 amplitudes as parameters for the genetic algorithm. In the calculations performed in this section, the number of individuals is kept fixed at 70 and the calculations were done with the same parameters used in Fig. 5 (main text).  \\
Fig. \ref{GA_complexity} shows the results obtained on three different version of cyanidin system. In the first we included in the calculation 20 excited states instead than 10. Because of the automatic way in which the frequencies are chosen, this corresponds to 15 frequencies, i.e. 45 amplitudes to be optimized instead than 39. Then, we scaled the 20 states in such a way to have the same energy range than in the 10 states calculation. In particular, we kept fixed the energies of the ground state and of the first excited state, while we scaled the others in such a way that the 20th excited state has the same energy of the 10th. This allows us to check what happens when the energy levels are closer the ones to the other and possibly other paths open up to go from the ground state to the target state. In this case the energy range is the same that the one for the 10 states calculations, and the amplitudes to be optimized are again 39.
Finally, to further increase the number of parameters to optimize, we performed an optimization calculation on the same 10 states cyanidin system but asking for 24 Fourier frequencies instead than the standard 12, as usual plus the one corresponding to i = 0. This means this time there are 75 parameters for the algorithm to optimize.
In Fig. \ref{GA_complexity} the cyan plot shows the results obtained on cyanidin including in the calculation 20 excited states instead than 10. Comparing the results obtained with the ones in Fig. 5, we can see that we have a decrease in the performance, with the value of the target state population after 100 iteration which is 0.955 instead than 0.989. For this reason we tried to improve this result adding 50 more convergence iterations and we were able to obtain a final population of 0.972, with similar results on the values of $J$ and field fluency. As expected more iterations, possibly focusing on both convergence and exploration, are able to solve the problem of a greater complexity.\\
We then compared this result on the 20 state system with the other two variations of the cyanidin system: in the case of 20 scaled states (solid pink line), we have a slightly worst result that in the not scaled case, even if the difference can easily be a consequence of performance fluctuations in different runs of genetic algorithm. The third case (solid brown line), the one with 25 frequencies for 10 excited states, performs similarly to the others.

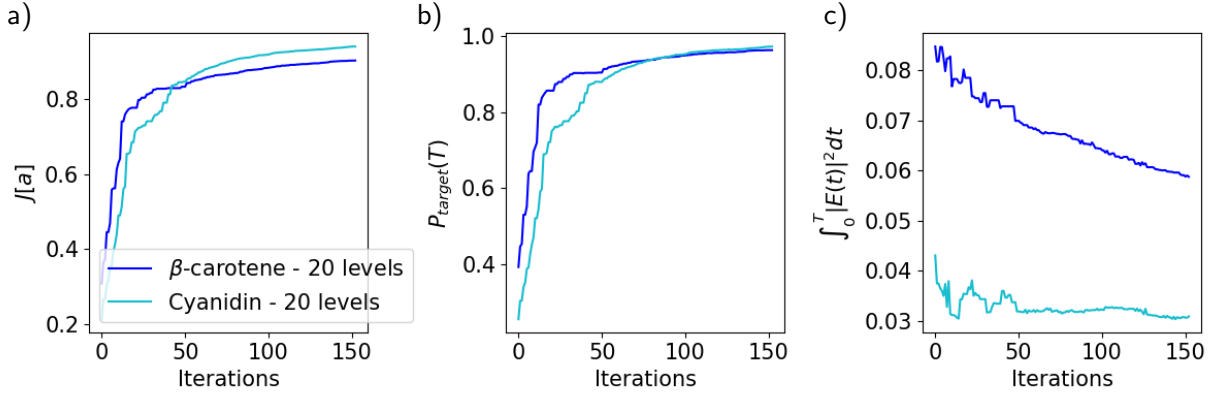
\begin{figure*}[h!]

\begin{tikzpicture}[node distance=cm,
    every node/.style={fill=white, font=\sffamily}]
    
\node (figure) at (0,0)    {\centering
    \includegraphics[width = \textwidth]{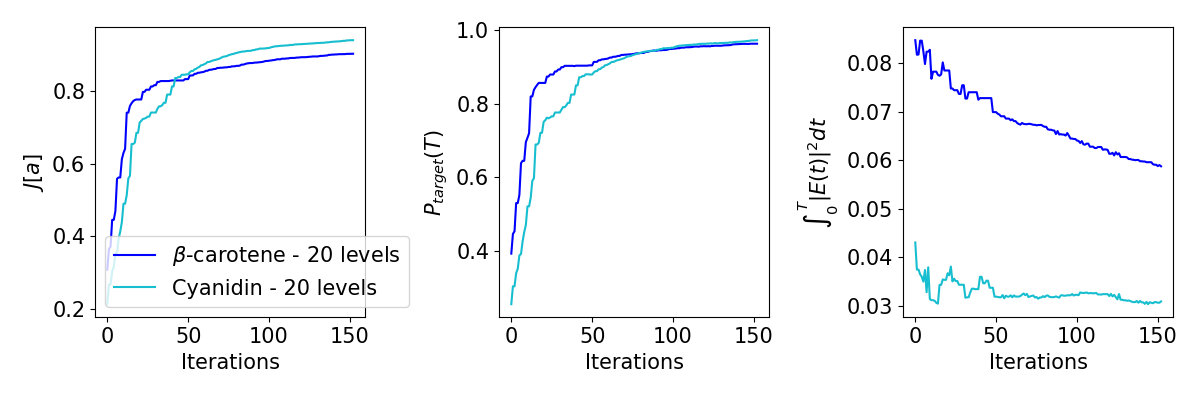}
};
    
\node (a) at (-7.85, 2.55) {a)};

\node (b) at (-2.4, 2.55) {b)};

\node (c) at (3.00, 2.55) {c)};

\end{tikzpicture}

\caption{\label{beta_carotene} Comparison of the optimal control problem solution between cyanidin and $\beta$-carotene. Plots of $J$ (a), target state ($|1\rangle$) population (b) and field fluency (c). Color code: cyanidin - solid cyan line; $\beta$-carotene - solid dark blue line.}

\end{figure*}

Finally, as our results has been all obtained on cyanidin, we tested our method on $\beta$-carotene, a molecule with a linear structure very different from the cyanidin aromatic one. We included in the description 20 energy states, which result in an energy range similar to the one of cyanidin. Fig. \ref{beta_carotene} shows how the results obtained with the genetic algorithm are very similar in term of performances with respect to the one obtained on the 20 state cyanidin system. 

\section*{APPENDIX C}

\begin{figure*}[htbp]

\begin{tikzpicture}[node distance=cm,
    every node/.style={fill=white, font=\sffamily}]
    
\node (figure) at (0,0) {\centering
    \includegraphics[scale = 0.45]{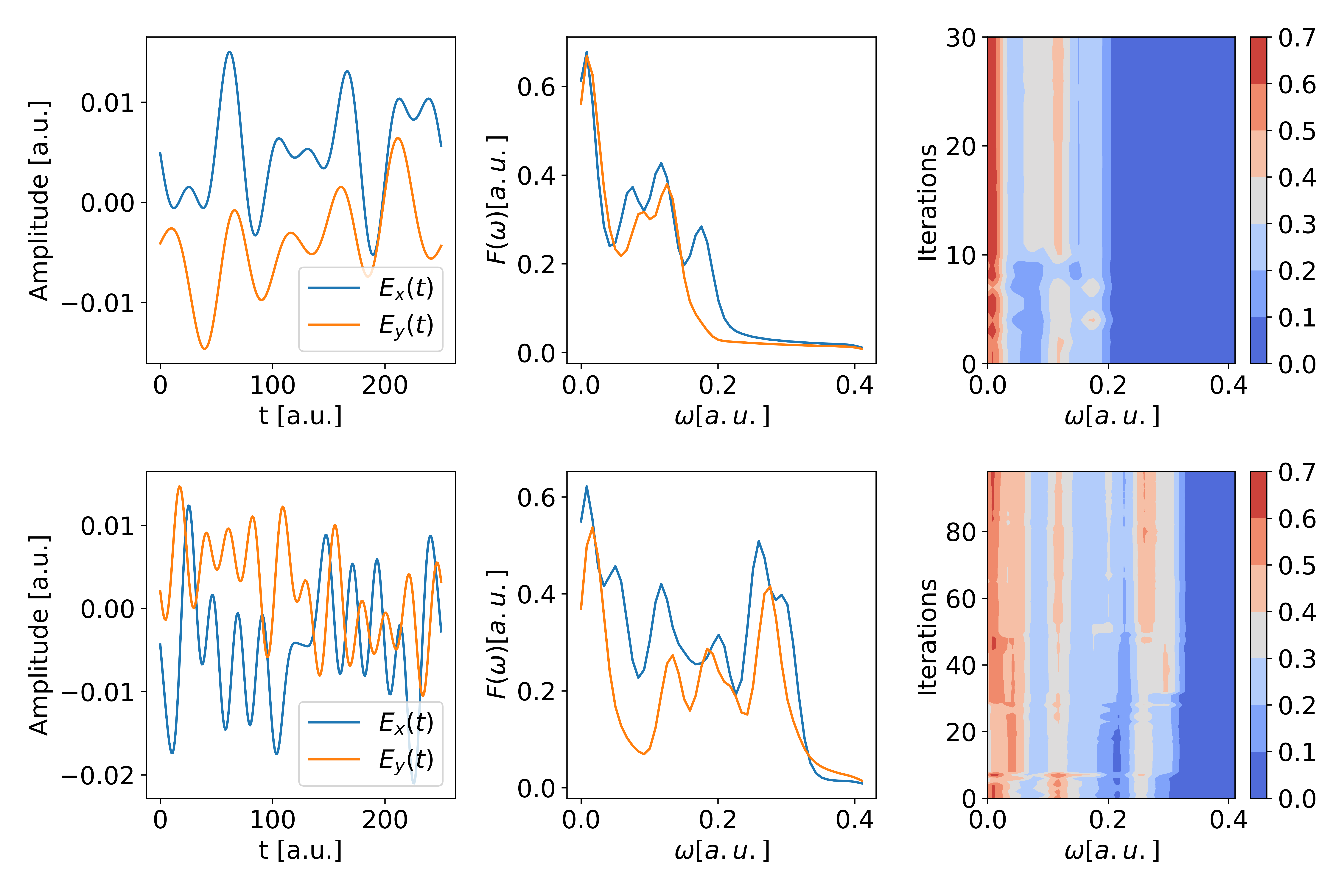}
    };
    
\node (system 2) at (-7.65, 2.5) {\textbf{Cyan3}};

\node (system 1) at (-7.65,-2.2) {\textbf{Cyan11}};

\node (a) at (-5.7, 4.25) {a)};

\node (b) at (-1.5, 4.25) {b)};

\node (c) at (2.50, 4.25) {c)};

\node (d) at (-5.7, 0) {d)};

\node (e) at (-1.5, 0) {e)};

\node (f) at (2.50, 0) {f)};

\end{tikzpicture}

\caption{\label{freq_analysis} Results of the optimal control problem solved with the TS-PCA propagation coupled with a GA. Upper row plots (a-c): optimal pulse (a), frequency distribution (b) and evolution of the frequency distribution (c) during the optimization for system \textbf{Cyan3}. Bottom row plots (d-f): optimal pulse (d), frequency distribution (e) and evolution of the frequency distribution (f) during the optimization for system \textbf{Cyan11}. Heatmaps show the frequency distributions of the \textit{x} component of the fields.}
\end{figure*}
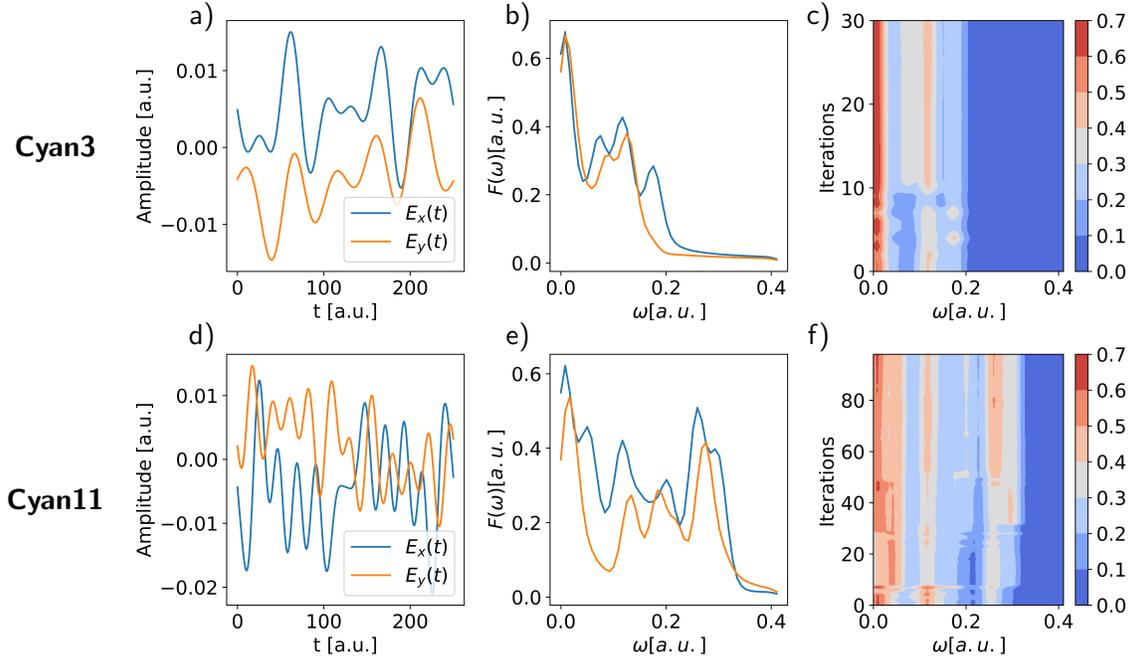

Here we present an analysis of the optimal control pulse obtained with the approach described in the main text. In Fig. \ref{freq_analysis} we reported the optimal result, i.e. the laser pulse after the optimization and its Fourier transform, obtained with the hybrid implementation using a GA optimizer for the control problem referring to system \textbf{Cyan11} (upper row Fig. \ref{freq_analysis}a-\ref{freq_analysis}b) and to system \textbf{Cyan3} (bottom row Fig. \ref{freq_analysis}d-\ref{freq_analysis}e); later we will discuss also panels \ref{freq_analysis}c-\ref{freq_analysis}f reporting the evolution of the field frequency distribution as a function of the iterations for the \textit{x} component of the field. We reported in both cases only the plots for the $x$ and $y$ components as, due to the planarity of our system, the transition dipole moment along the $z$ direction is almost null. As we already mentioned in describing the purpose of optimal control, the optimal field not only realizes the direct transition from the ground to the target state, but must also discourage subsequent and competitive transitions to other excited states \cite{klamroth:2006,rosa:2019}. This means that, even if the direct transition from the ground state to the target state corresponds to $\omega =$ 0.126\,a.u., other frequencies are present in the optimal field. Looking at the plots, it is evident how increasing the complexity of the system leads to an increase in the complexity of the optimal field. \textbf{Cyan11} system has a larger number of electronic states with respect to \textbf{Cyan3}, which also means a greater number of possible path to reach the optimal solution. Indeed, we can see that the frequency distributions reveal some contributions which are even more intense than the one corresponding to the direct transition from the ground state to the target state ($\omega =$ 0.126\,a.u.),  to be ascribed either to multistep paths among different states or to processes of suppression of competitive transitions. This is an established result of solving optimal control problems which the genetic algorithm is able to reproduce.
Nevertheless, there is also a second reason for the larger number of frequencies in the larger system which is a consequence of how the field is represented in the case of the GA optimizer: in the case of \textbf{Cyan11} the number of harmonics, and consequently amplitudes, which are summed up to describe the field is larger with respect to system \textbf{Cyan3}. In principle {\it useless} amplitudes, i.e. the ones which do not have a direct effect on the final value of the population in the target state, should be minimized and their value should be set to zero. In practice the GA optimizer is more efficient in identifying, and optimizing, amplitudes which have a direct effect on the result (i.e. amplitudes which favor or discourage the transition). 

As a consequence, it is possible that some of the smaller amplitudes are a consequence of an incomplete minimization. Nevertheless, the second term in the $J$ functional (i.e. the integral accounting for the filed fluency) works exactly in the way of minimizing those {\it useless} amplitudes, which means that if needed, a finer tuning of the optimization algorithm would lead to a further improved result. 
Now, we discuss panels \ref{freq_analysis}c-\ref{freq_analysis}f which show frequency distribution of the fields as a function of the iterations for direction \textit{x}. With regard to \textbf{Cyan3} system (Fig. \ref{freq_analysis}c) we can see that all along the optimization the main contribution to the field is given by a low lying frequency component which, as we are dealing with a very simple system, we can easily assign to the frequency resonant with the transition between the first two excited states; in this case it is easy to rationalize the role of this frequency component as it allows to bring back to the target state all the population which has been mistransferred to the second excited state. On the other hand, in the case of system \textbf{Cyan11}, the heatmap of Fig. \ref{freq_analysis}f allows to appreciate even more the main information that we have extracted from the frequency distribution of the optimal result: to achieve high controllability within a manifold of electronic states in ultrashort time scales, alternative paths including transitions are important as much as the contribution due to the frequency resonant with the transition of interest. This general statement is in agreement with other analysis carried out in literature \cite{klamroth:2006,rosa:2019}. An in-depth analysis of the population dynamics and the fluxes among excited states is beyond the purpose of this work and will be carried out in a future study.


\section*{APPENDIX D}

\begin{figure}[!htbp]
    \centering
    \includegraphics[width = \textwidth]{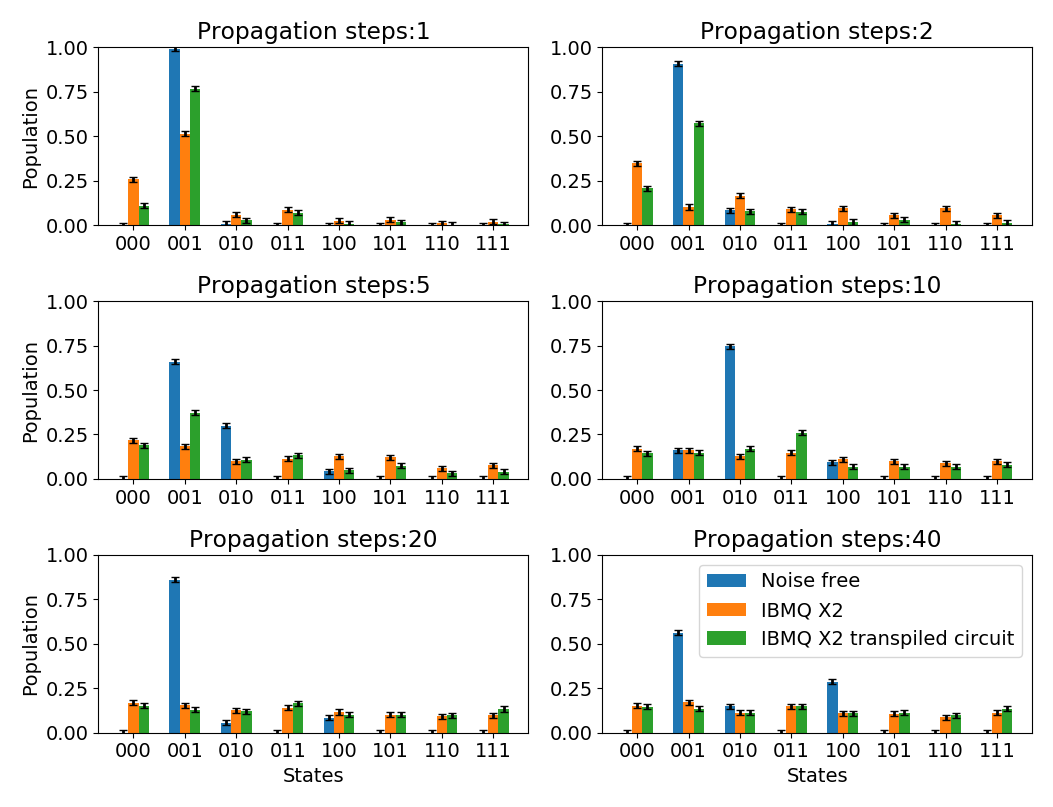}
    \caption{Quantum simulation of system \textbf{Cyan3}. Experimental probability distributions obtained for the quantum simulation circuit (second row of Table I, main text) on the IBM Q X2 hardware; each panel shows the results obtained performing different propagation steps, in blue the reference noise free distribution. We recall that at t=0 the molecule is in the ground state $|001\rangle$. Error bars represent the standard error computed assuming Gaussian statistics for 2048 repetitions. }
    \label{population_quantum_hardware}
\end{figure}

Here we show additional results for the quantum simulation performed with the mapping discussed in Sec. II.A (main text). Probability distributions obtained measuring the computational basis states population (Fig. \ref{population_quantum_hardware}) confirm the considerations outlined discussing the results of the fidelity plots (main text). We can appreciate how the presence of noise drives the evolution of the quantum computer out of the subspace of the bitstring states with only one qubit in state $|1\rangle$ until it reaches thermalisation with all the states equally populated. 

\newpage

\bibliography{references}

\end{document}